\documentclass[pre,two column,notitlepage]{revtex4-1}
\usepackage[utf8]{inputenc}

\date{\today}

\usepackage{graphicx}
\usepackage{csquotes}
\usepackage{amsmath}
\usepackage{amssymb}
\usepackage{comment}
\usepackage{xcolor}
\usepackage[caption=false]{subfig}

\newcommand{\e}{\mathcal{E}}

\newcommand{\sech}{\text{sech}}
\newcommand{\Omegai}{\Omega_i}

\newcommand{\Omegaid}{\Omega_{i\zeta_i}}

\begin{document}
\title{Steady states and coarsening in one-dimensional driven Allen-Cahn system}
\author{Rajiv G. Pereira}
\affiliation{School of Physics, IISER Thiruvananthapuram, Vithura, Kerala 695551, India}

\begin{abstract}

We study the steady states and the coarsening dynamics in a one dimensional driven non-conserved system modelled by the so called driven Allen-Cahn equation, which is the standard Allen-Cahn equation with an additional driving force. In particular, we derive equations of motion for the phase boundaries in a phase ordering system obeying this equation using a nearest neighbour interaction approach. Using the equations of motion we explore kink binary and ternary interactions and analyze how the average domain size scale with respect to time. Further, we employ numerical techniques to perform a bifurcation analysis of the one-period stationary solutions  of the equation. We then investigate the linear stability of the two-period solutions and thereby identify and study various coarsening modes.  
\end{abstract}

\maketitle

\section{Introduction}

 Phase ordering in systems quenched suddenly from a disordered to an ordered phase is ubiquitous in nature. It is manifest not only in physical phenomena such as domain growth in ferromagnets and spinoidal decomposition in alloys~\citep{Cahn_Spinoidal, Christiansen_2020, MULLER2015181}, but also in other natural processes such as the formation of membraneless cell compartments~\cite{Alberti_2019}, and the spontaneous organization of unicellular organisms like bacteria and amoeba~\cite{Liu_2019}.  Not surprisingly, it has been a topic of extensive research in the last several decades~\citep{Bray_review, onuki2002phase, Almeida2021} and still continues to intrigue scientists across various disciplines~\citep{Gasior_2019, Berry_2018}. Over the years, this phenomenon has been studied in experiments, simulations~\citep{Bouttes2014, Arenzon_2015, Holyst_2001, Dai_2016, BENES2004187}, and by means of various numerical and analytical techniques~\citep{Linden_2019, Nepomnyashchy_2015, lifshitz1962, Lifshitz1961}.

Mesoscopic descriptions of domain coarsening in terms of a coarse grained order parameter have been successful in explaining numerous interesting features~\citep{cahn1977microscopic, Cahn_Spinoidal, KAWASAKI1982573, Hulsmann_2021, Nepomnyashchy_2015}. Two such mathematical models that are predominant in literature are given by the Cahn-Hilliard (CH) and  the Allen-Cahn (AC)  equations~\citep{Bray_review, CH}. These are equilibrium models that can be derived from the Landau-Ginzburg free energy~\citep{HalperinClassic}. The CH equation follows conserved dynamics, 
while the AC equation follows non-conserved dynamics. These equations and their various extensions have been studied in a variety of contexts~\citep{Hulsmann_2021, Ren_2016, John2005}.

In Ref.~\citep{leung1990theory}, Leung studied phase separation in a driven conserved lattice gas by introducing a driving term to the CH equation.
The resulting equation is referred to as the convective Cahn-Hilliard (cCH) equation. It has also been used to describe several other physical processes like spinoidal decomposition in the presence of gravitational field~\citep{EmmotBray} and faceting in crystals~\citep{Golovin_1998, Golovin_1999}. Various aspects of the cCH equation have been studied in detail over the years. Coarsening mechanisms such as kink binary and ternary coalescence, scaling of the average domain size with respect to time, and bifurcation analysis of the stationary and the travelling wave solutions are some of them.  These studies have revealed several interesting features of the cCH equation and have helped us understand coarsening in driven conserved systems~\citep{PODOLNY2005291,Golovin2001, EmmotBray, Watson_2003, Tseluiko_2020}.    

However, the effects of a driving force on the coarsening dynamics of a non-conserved system, for instance, a  driven non-conserved lattice gas, are comparatively less explored. Such a system can be aptly modelled by adding a similar driving term to the AC equation. The resulting equation, which we refer to as the driven Allen-Cahn (dAC) equation, is the focus of this paper. We note here that the critical dynamics of a stochastic version of the dAC equation was explored in Ref.~\citep{bassler_1994}. 

 Several interesting questions naturally arise here. What are the steady state solutions of the dAC equation and how do they differ from those of the AC and the cCH equations?  How does the average domain size scale with respect to time in the presence of the drive when the dynamics is non-conserving? What are the allowed coarsening mechanisms? How are they affected by the strength of the drive and how do they differ from those of the conserved model?   

Motivated by these questions, we employ analytical and numerical techniques to study the steady state solutions and the coarsening dynamics of the dAC equation. In particular, we first study single (anti)kink steady state solutions in an infinite domain using asymptotic analysis. Secondly, we use analytical techniques to derive equations of motion for the phase boundaries at large driving strength in a system with multiple kinks and antikinks separated by large distances. The equations of motion are then exploited to investigate various coarsening mechanisms and to obtain the scaling form for the average domain size with respect to time. We also perform simulations of kink binary coalescence and compare them with the analytical results. Thirdly, we do a bifurcation analysis of the one-period stationary solutions of the dAC equation with the help of the continuation and bifurcation software Auto07p~\citep{Auto07p}. Lastly, a linear stability analysis of the two-period solutions is performed to obtain the coarsening modes and the corresponding eigenvalues. We also explore the behavior of the modes and the eigenvalues as a function of the domain size $L$ and the driving strength $\e$.  

This paper is organized as follows. In Sec.~\ref{asymptotic}, we present the asymptotic analysis of the single (anti)kink steady state solutions. Section~\ref{sec:nneighbour} is dedicated to the derivation of the equations of motion for the phase boundaries and the simulations, and Sec.~\ref{sec:oneperiod} to the bifurcation analysis of the one-period stationary solutions and the linear stability analysis of the two-period solutions.

\section{Single (anti)kink solutions} \label{asymptotic}
In this section, we first introduce the dAC equation and then study the fixed points of its time independent version in frame moving with a constant velocity $v.$ We also establish that single (anti)kink travelling wave solutions exist only for isolated values of $v$ using asymptotic analysis. 

The dAC equation is obtained by adding a driving force proportional to $\psi \psi_x$  to the standard AC equation and is explicitly written, in dimension less form, as
\begin{equation}\label{cAC}
    \psi_t=\psi_{xx}+\psi- \psi^3 + 2\e \psi \psi_x,
\end{equation}
where $\e$ is the driving strength and $\psi(x,t)$ is the order parameter. The subscripts $x$ and $t$ denote derivatives with respect to the spatial coordinate $x$ and time $t$, respectively. In the limit $\e \rightarrow 0$, we retrieve the AC equation. Note that the driving term $2\e \psi \psi_x$ is of non-equilibrium nature as it cannot be derived from a Hamiltonian. This is in contrast to the AC equation, where the forces can be derived from the Landau-Ginzburg Hamiltonian~\citep{HalperinClassic, Bray_review}.  



 For the following analysis, it is convenient to rewrite Eq.~(\ref{cAC}) in a frame moving with a constant velocity $v$.  
    \begin{equation}\label{dACv1d}
        \psi_t=v \psi_x +\psi_{xx} + \psi -\psi^3 +2 \e \psi \psi_x.   
    \end{equation}
Clearly, the time independent solutions of Eq.~(\ref{dACv1d}) correspond to the travelling wave solutions of the dAC equation in the rest frame. Note that both Eq.~(\ref{dACv1d}) and Eq.~(\ref{cAC}) are symmetric under the transformation $(\mathcal{E}, \psi) \rightarrow (-\mathcal{E}, -\psi)$. Therefore we consider only values of $\mathcal{E} \ge 0$ throughout this paper as it is sufficient.  

We first examine the linear stability of the constant solutions $\psi_0$ of Eq.~(\ref{dACv1d}). Unlike the cCH equation for which any constant function is a solution, the dAC equation has only three spatially uniform time independent solutions, namely $\psi_0=0,\pm1$. The linear stability of these solutions can be investigated by perturbing them as
 \begin{equation}
     \psi(x, t)=\psi_0 + \epsilon \exp{(\beta t + i k x)},
 \end{equation}
 where $\epsilon$ is a small parameter. Substituting this in Eq.(\ref{dACv1d}) and subsequent linearization yields the following dispersion relation. 
 \begin{equation}\label{betaofk}
    \beta(k)=-k^2+(1-3 \psi_0^2) + ik\, (v+2 \mathcal{E} \psi_0).
\end{equation}
The sign of the real part of $\beta$ determines the linear stability of the solutions, where
\begin{equation}\label{betareal}
    \text{Re}(\beta)=
    \begin{cases}
    -k^2-2, & \psi_0=\pm1 \\
    -k^2+1, & \psi_0=0.
    \end{cases}
\end{equation}
It follows that in an infinite domain $\psi_0=\pm1$ are stable, and $\psi_0=0$ is unstable with respect to time dependent perturbations.   

We now analyse the fixed points of the time independent form of Eq.~(\ref{dACv1d}) 
     \begin{equation}\label{dACv1dtind}
     v \psi_x +\psi_{xx} + \psi -\psi^3 +2 \e \psi \psi_x=0.      
    \end{equation}
Clearly, the fixed points of the above equation are also given by $\psi_0=0,\pm1$. Substituting $\psi=\psi_0+\epsilon \exp{h x}$ in Eq.~(\ref{dACv1dtind}) and linearizing the resulting equation in the small parameter $\epsilon$, we obtain the following eigenvalue equation.
\begin{equation}\label{evaleq}
    h^2+(v+2\e \psi_0) \; h-(3 \psi_0^2-1)=0.
\end{equation}
Of course, Eq~(\ref{evaleq}) could have been obtained by setting $ik=h$ and $\beta=0$ in Eq.~(\ref{betaofk}) as well. The eigenvalue equation~(\ref{evaleq}) is readily solved to obtain the following three cases: 
\begin{equation}\label{evals}
h=\frac{1}{2} \times
    \begin{cases}
       -(v+2 \e) \pm \sqrt{(v+2 \e)^2+8}, & \psi_0 =1, \\
      -v \pm \sqrt{v^2-4}, & \psi_0=0, \\
      -(v-2 \e) \pm \sqrt{(v-2 \e)^2+8}, & \psi_0=-1.
    \end{cases}
\end{equation}
From the above solution, we deduce that for $\psi_0=\pm 1$ there are two eigenvalues, one negative and one positive. Therefore for the fixed points $\psi_0=\pm1$ the dimensions of the stable manifolds $W^S(\psi_0=\pm1)$ and the unstable manifolds $W^U(\psi_0=\pm1)$ are both $1$. A stable single kink solution $\Omega_+$ should be such that $\Omega_+(\pm \infty)=\pm1$. Note that $\psi_0=0$ is unstable with respect to time dependent perturbations in an infinite domain. It follows that a kink should lie in the intersection 

\begin{equation}
    W^U(\psi_0=-1) \cap W^S(\psi_0=1).
\end{equation}
Similarly a stable single antikink solution $\Omega_-$ should lie in the intersection
\begin{equation}
    W^U(\psi_0=1) \cap W^S(\psi_0=-1).
\end{equation}
These are both intersections of two one dimensional manifolds in a two dimensional space. This restricts the choice of the parameter $v$. Therefore we expect a kink to exist only for isolated values of $v$. This is in contrast to the case of the cCH equation where kink solutions exist for values in a region in that parameter space, as has been shown using a similar analysis in previous works~\citep{Golovin_1998, EmmotBray, PODOLNY2005291}. However, antikink solution exist only for isolated points in the parameter space in the case of the cCH equation as well. 

The following solutions corresponding to $v=0$ exist for Eq.~(\ref{dACv1dtind}): 
\begin{equation}\label{tanhsoln}
    \Omega_\pm (x)=\pm \tanh{s_\pm x},
\end{equation}
where
\begin{equation}\label{sis}
 s_\pm=  (1/2) \sqrt{2+\e^2} \pm \e/2   
\end{equation}
The signs $+$ and $-$ correspond to kink and antikink, respectively. 

The following remark on the nature of the fixed point $\psi_0=0$ is in order here. We deduce from Eq.~(\ref{evals}) that for the fixed point $\psi_0=0$, the corresponding eigenvalues are $h= \pm \sqrt{-4}$ when  $v=0$.  These are pure imaginary numbers, implying that the fixed point $\psi_0=0$ is a center. Therefore we expect there to be periodic orbits in the neighbourhood of this fixed point in the phase-plane. We shall  explicitly show in Sec.~\ref{sec:oneperiod} that there indeed exist periodic solutions of Eq.~(\ref{dACv1dtind}) that oscillate about $\psi_0=0$ when $v=0.$ 

We now proceed to examine the motion of phase boundaries in a system with multiple kinks and antikinks.

\section{Nearest neighbour interaction theory for phase boundaries}\label{sec:nneighbour}

In this section, we first analytically study the coarsening process in a driven one dimensional Allen-Cahn system with multiple phase boundaries. To this end, we derive equations of motion governing the dynamics of the phase boundaries using a nearest-neighbor interaction approach. Using the equations of motion, we examine kink binary and ternary interactions and obtain a scaling law for the average domain size with respect to time. We then simulate kink binary coalescence with the help of Mathematica and compare them with the analytical results. 

\subsection{Equations of motion}

To derive the equations of motion, we first write down an ansatz to the dAC equation that has a series of alternating kinks and antikinks, which are separated by large domains where the order parameter is more or less constant. We then substitute it in the dAC equation and concentrate on the positions of the phase boundaries. 

Remember that the constant solutions of the dAC equation that are stable with respect to time dependent perturbations are given by $\psi_0=\pm1$. Hence, we expect that in a region far left (right) to a kink (antikink) $\psi \simeq -1$, and likewise in a region far right (left) to a kink (antikink) $\psi \simeq 1$. Such a (anti)kink solution is already known from the previous section [see Eq.~(\ref{tanhsoln})]. We adopt these $\tanh$ profiles and construct the ansatz as described below. 

Let $p_i(t)$ denote the position of the $i$th phase boundary. Without loss of generality we assume that odd $i$ correspond to kinks and even $i$ correspond to antikinks, and for convenience we introduce the following set of co-moving coordinates.
\begin{equation}\label{zeta}
    \zeta_i=x-p_i(t).
\end{equation}
\begin{figure}
    \centering
    \includegraphics[scale=0.8]{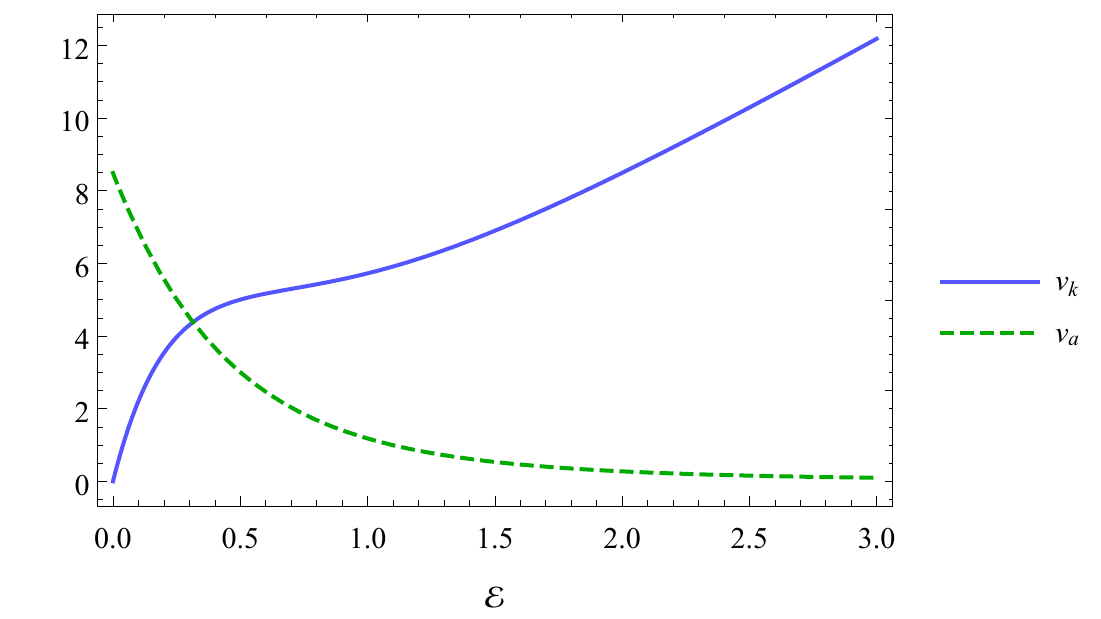}
    \caption{The coefficients $v_k$ and $v_a$ vs the driving strength $\e$. The two lines intersect at $\e=0.3$.}
    \label{fig:vkva}
\end{figure}
The solution near the $i$th phase boundary is constructed as a moving $\tanh$ profile superposed with a small correction:
\begin{equation}\label{ansatz}
    \psi_i(x,t)=\Omega_i(\zeta_i)+\omega_i,
\end{equation}
where
\begin{align}\label{CapitalOmega}
    \Omega_i(\zeta_i)=
    \begin{cases}
       \Omega_+(\zeta_i), & \text{odd $i$}, \\
       \Omega_-(\zeta_i), & \text{even $i$}, 
    \end{cases}
\end{align}
and $\omega_i$ is the small correction. We write $\omega_i$ as a sum of contributions from the immediate neighbours:
\begin{equation}\label{omegasplit}
 \omega_i=  \omega_i^- + \omega_i^+,   
\end{equation}
where
\begin{align}\label{smallomega}
    \omega_i^\mp= &\Omega_{i\mp1}(\zeta_{i\mp1})-\Omega_{i\mp1}(\pm \infty).
    \end{align}
As the distance between the phase boundaries is large, $\omega$ is indeed small near $p_i$. A similar nearest-neighbour interaction approach has been used previously in the case of the standard AC and CH equations~\citep{KAWASAKI1982573, PODOLNY2005291}.  For compactness, we introduce the force
\begin{equation}\label{FF}
    F[\psi,\psi_x] \equiv \psi - \psi^3 + 2 \e \psi \psi_x.
\end{equation}
The dAC equation is written in terms of $F$ as
\begin{equation}\label{compactdAC}
        \psi_t= \psi_{xx}+F[\psi,\psi_x].
\end{equation}
It will be handy to note here that $\Omega_i$ and $\Omegaid$ satisfy the relations
\begin{equation}\label{omegaeq}
    \Omega_{i\zeta_i{\zeta_i}}+ F[\Omega_i, \Omega_{i\zeta_i}]=0,
\end{equation}
and
\begin{equation}\label{omega'}
\! \! \!    \left[\frac{\partial^2}{\partial \zeta_i^2}+ \frac{\partial F[\Omega_i,\Omega_{i\zeta_i}]}{ \partial \Omega_{i\zeta_i}} \frac{\partial}{\partial \zeta_i} + \frac{\partial F[\Omega_i,\Omega_{i\zeta_i}]}{\partial \Omega_i} \right]\Omega_{i\zeta_i}=0,
\end{equation}
respectively. Substituting Eq.~(\ref{ansatz}) in Eq.~(\ref{compactdAC}), and then using Eq.~(\ref{omegaeq}) and (\ref{smallomega}) we obtain
\begin{align}\label{inbetween1}
\! \! \!     \!\sum_{j=-1}^1 \! \!-\dot{p}_{i+j} \Omega_{i+jx}=& \omega_{xx} \! +\!\frac{\partial F[\Omega_i,\Omega_{i\zeta_i}]}{\partial \Omega_i} \, \omega -2 \e (\Omega_i \omega)_x +\widetilde{F}_i,
\end{align}
where 
\begin{equation}\label{Ftilde}
    \widetilde{F}_i \equiv F[\psi_i,\psi_{ix}]-F[\Omega_i,\Omega_{i\zeta_i}]-\frac{\partial F[\Omega_i,\Omega_{i\zeta_i}]}{\partial \Omega_i} \, \omega+2\e (\Omega_i \omega)_x,
\end{equation}
and the dot above symbols represents derivative with respect to time. 

Acting the operator $\int_{-\infty}^{\infty}\!dx \,\Omega_{ix}$ on Eq.~(\ref{inbetween1}) and then using Eq.~(\ref{omega'}) in the right hand side after integration by parts, we obtain
\begin{equation}\label{nnnoapprox}
   \sum_{j=-1}^1 \!-\dot{p}_{i+j}\int_{-\infty}^{\infty} \! \! dx \, \Omega_{ix} \Omega_{i+jx}  =\int_{-\infty}^{\infty} \! dx \, \Omega_{ix}  \widetilde{F}_i.
\end{equation}
The left hand side (LHS) and the right hand side (RHS) of the above equation can be simplified by approximating the integrals therein using Eqs.~(\ref{approx}) and (\ref{approx2}). These approximations are valid when the driving strength $\e$ as well as the separation 
\begin{equation}\label{li}
l_i=p_{i+1}-p_i,    
\end{equation}
 are large. See Appendix~\ref{appendix:integrals} for further details. After the straight forward calculations shown therein, we obtain the following equations of motion for the cases of odd and even $i$, respectively. 
\begin{widetext}
\begin{align}\label{eom1}
     -\frac{1}{3}s_+ \dot{p}_i
    + 2 s_- \dot{p}_{i+1} e^{-2 s_- l_i} 
     &+ 2 s_- \dot{p}_{i-1} e^{-2 s_- l_{i-1}} \nonumber \\
     &=4 \e \left( \frac{s_+}{3}+2 s_- \frac{s_-}{s_+} \right)e^{-2 s_- l_{i-1}}
    -4 \e \left( \frac{s_+}{3}+2 s_- \frac{s_-}{s_+} \right)e^{-2 s_- l_{i}}, \;& \text{(odd $i$),} \\
 -\frac{1}{3}s_- \dot{p}_i+2 s_- \dot{p}_{i+1}e^{-2 s_-l_i} &+2 s_- \dot{p}_{i-1}e^{-2 s_-l_{i-1}} \nonumber \\
 &=2\left\{1-2 \e s_-\left(1+\frac{s_-}{s_+}\right) \right\}e^{-2s_-l_{i-1}}
    - 2\left\{1-2 \e s_-\left(1+\frac{s_-}{s_+}\right) \right\}e^{-2 s_-l_i}, \; &\text{(even $i$).}\label{eom2}
\end{align}
\end{widetext}

\subsection{Kink binary interaction and scaling law.}

Consider a kink binary consisting of a kink and an antikink that are closer to each other than to the other respective adjacent phase boundaries. Let $p_1$ and $p_2$ be their respective positions. Substituting $p_0$, $p_1$, and $p_2$ for $p_{i-1}$, $p_i$, and $p_{i+1}$, respectively, in Eq.~(\ref{eom1}) and  $p_1$, $p_2$, and $p_3$ for $p_{i-1}$, $p_i$, and $p_{i+1}$, respectively, in Eq.~(\ref{eom2}) and then solving the resulting equations simultaneously leads to the equations of motion  
\begin{align}\label{binaryeom}
    &\dot{p}_1=v_k \, e^{-2 s_- l_1},
    &\dot{p}_2=-v_a \, e^{-2 s_- l_1},
\end{align}
where
\begin{align}\label{vkva}
    v_k&=12 \e \left[\frac{1}{3}+ 2 \left(\frac{s_-}{s_+}\right)^2
    \right], \nonumber \\
    v_a&=\left[\frac{6}{s_-} - 12 \e \left( 1+\frac{s_-}{s_+}\right) \right],
\end{align}
and $l_i$ is given by Eq.~(\ref{li}). Note that terms with $\exp(-2s_-l_0)$ and $\exp(-2 s_- l_2)$ are discarded as ${l_0, l_2>>l_1}$. The coefficients $v_k$ and $v_a$ are plotted against the driving strength $\e$ in Fig.~\ref{fig:vkva}.

For all values of $\e>0$, $v_k$ and $v_a$ are positive. This implies that the kink at $p_1$ moves in the positive $x$ direction and the antikink at $p_2$ moves in the negative $x$ direction as is evident from Eq.~(\ref{binaryeom}). In other words, the kink and the antikink attract each other resulting in binary coalescence. We shall indeed see in Sec.~\ref{sec:oneperiod} that the dominant coarsening mode is the one where adjacent kinks and antikinks attract each other. 
This is in stark contrast to the case of the cCH equation where kink binary coalescence is impossible~\citep{Watson_2003}.

When $\e>0.3$, then $v_k>v_a$ indicating that the kink moves at a higher speed than the antikink. As $\e$ increases further, $v_k$ increases and $v_a \rightarrow 0$. It then follows from Eq.~(\ref{binaryeom}) that at large values of $\e$ the antikink is almost stationary, and the kink speed increases with $\e$ when $l_1$ is fixed.  These results agree with the results of the simulations explained at the end of this section. It is worth mentioning here that for the case of the standard Allen-Cahn equation the kink and the antikink moves with the same speed~\citep{KAWASAKI1982573}. 

When $\e<0.3$, then $v_k<v_a$ as can be seen from Fig~\ref{fig:vkva}. It then follows from Eq.~(\ref{binaryeom}) that the theory developed here predicts that the antikink moves at a higher speed than the kink. But this is not in agreement with the results of the simulations, which shows that the kink moves at a higher speed when $\e<0.3$ as well. However, the equations of motion derived in this section are not expected to hold for small values of $\e$ since the approximation we used, namely Eq.~(\ref{approx}), is not valid for small~$\e$. 

\begin{figure*}
\subfloat[$\e=0.1$, $l_1=10$, $t_1=30000$, $t_2=35000$, and $t_3=35474$. \label{fig:simu0p1}]{
\includegraphics[scale=0.45]{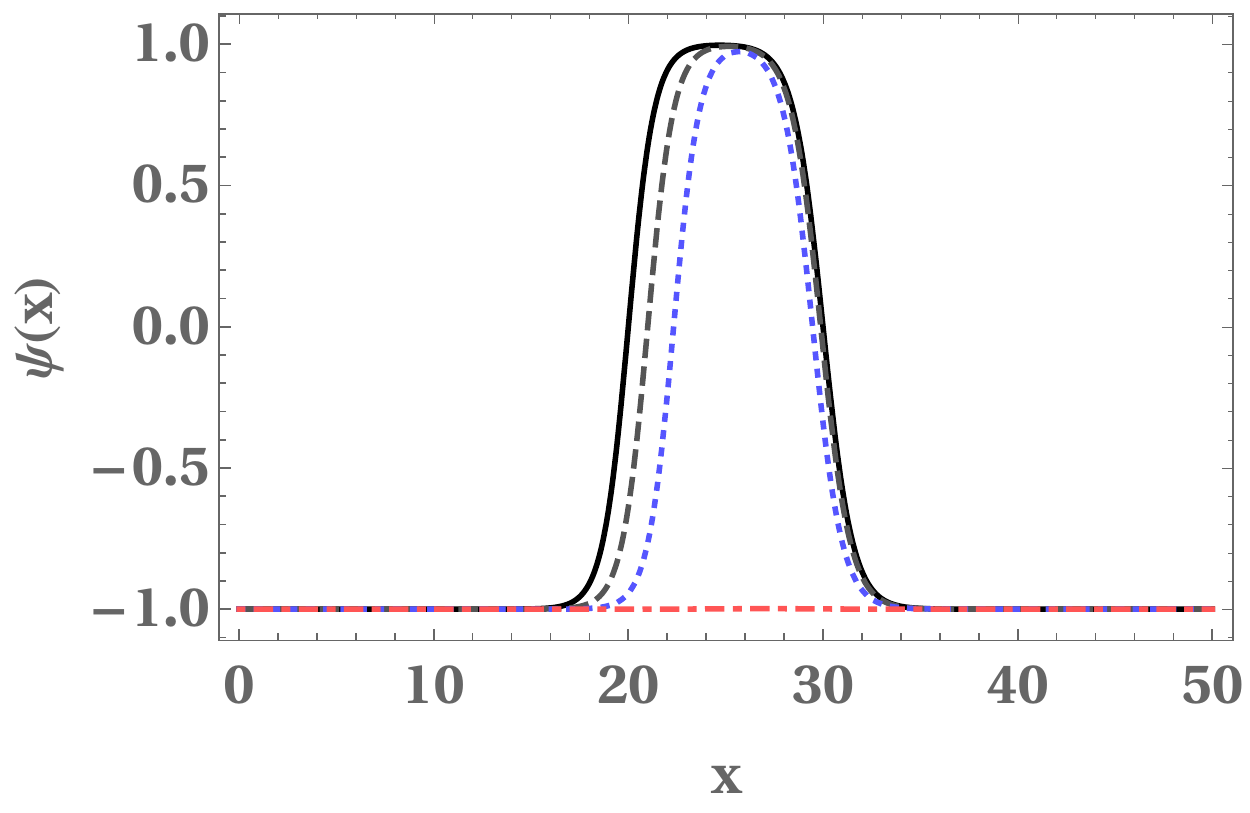}}
\subfloat[$\e=1$, $l_1=15$, $t_1=12000$, $t_2=15000$, and $t_3=15537$.\label{fig:simu1}]{
\includegraphics[scale=0.45]{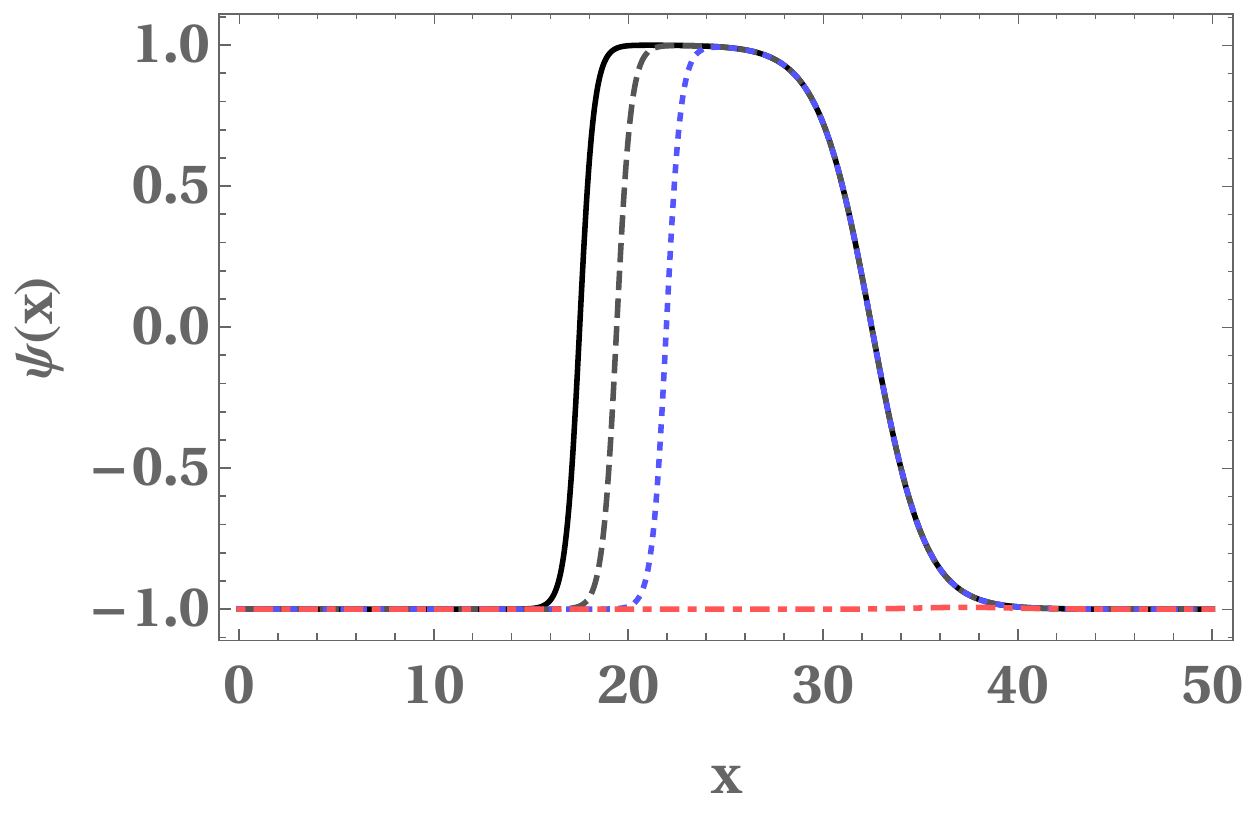}} 
\subfloat[$\e=1.5$, $l_1=15$, $t_1=1000$, $t_2=1350$, and $t_3=1457$. \label{fig:simu1p5}]{
\includegraphics[scale=0.45]{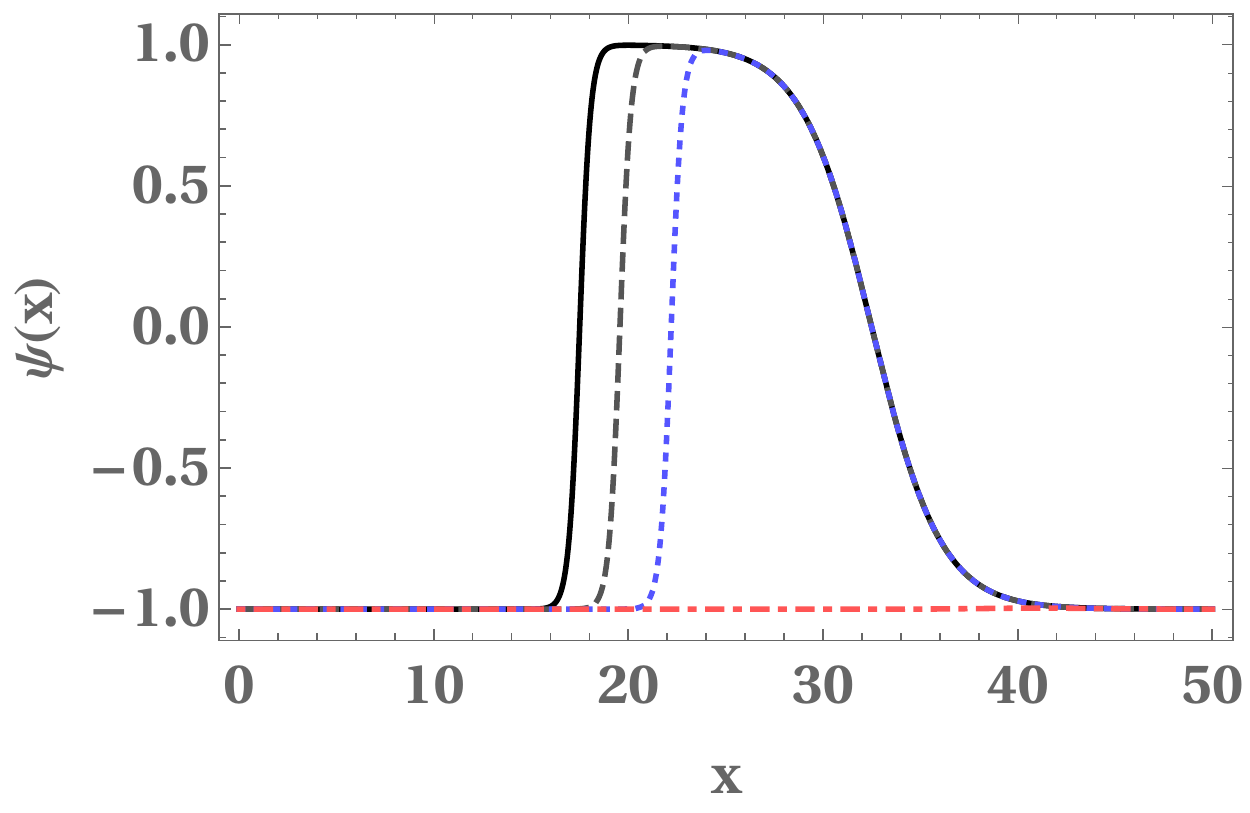}}
\caption{Time evolution of kink binary for different values of the driving strength $\e$. The black solid line shows the initial profile at $t=0$. The gray dashed, the blue dotted, and the red dot-dashed lines show the profiles at instants $t_1, \; t_2, \; \text{and} \; t_3$, respectively, where $t_1<t_2<t_3$. The instant $t_3$ is also the time at which the profile just becomes flat. The initial separation between the kink and the antikink is denoted by $l_1$.}
\label{fig:simu}
\end{figure*}
A dynamical equation for the separation $l_1$ can be easily obtained from Eq.~(\ref{binaryeom}) as shown below:
\begin{equation}
    \dot{l}_1=\dot{p}_2 - \dot{p}_1=-(v_k+v_a)e^{-2s_-l_1},
\end{equation}
which is readily solved to yield
\begin{equation}\label{domainlength}
    l_1 (t)=\frac{1}{2 s_-}\log\left[-\mu t + e^{2 s_- l_1(0)} \right],
\end{equation}
where
\begin{equation}
    \mu=2 s_- (v_k+v_a),
\end{equation}
and $l_1(0)$ is the initial distance between the kink and the antikink. From Eq.~(\ref{domainlength}) the time taken for a kink binary to coalesce can be readily calculated. Let $t_c$ be the time at which the binary coalesce. Then, by substituting $l_1(t_c)=0$ in Eq.~(\ref{domainlength}) we obtain
\begin{equation}\label{coalescence_time}
    t_c=\frac{e^{2 s_- l(0)}-1}{\mu}.
\end{equation}
Now consider a dAC system where the average separation between the kink binaries is given by $\Bar{l}$. Assume that $\Bar{l}$ as well as the driving strength $\e$ are large enough that the analytical results derived above can be applied. As the system evolves in time kink binaries coalesce, and with each such event $\Bar{l}$ increases by a fraction. Since the time taken for an average binary to coalesce grows exponentially with $\Bar{l}$ according to Eq.~(\ref{coalescence_time}), we in turn expect the average domain size $\Bar{l}$ to scale logarithmically with respect to time as also observed in simulation:
\begin{equation}
    \Bar{l}(t) \sim \ln{t}.
\end{equation}
Logarithmically slow coarsening is also observed in the case of the cCH equation when the characteristic length $\mathcal{L}$ is sufficiently larger than the \textit{Peclet} length $\mathcal{L}_P$~\citep{Watson_2003}. When $\mathcal{L}<<\mathcal{L_P}$, domain coarsening therein exhibits the coarsening rate $\mathcal{L}(t) \sim t^{1/2}$. 
\subsection{Kink ternary interaction}
We now examine kink ternary interaction. Two types of ternaries are possible in a one dimensional system with alternating kinks and antikinks: one where there is a kink in the middle of two antikinks and one where there is an antikink in the middle of two kinks. We call the former type-K ternary and the latter type-A ternary.

Consider a ternary of type-K. Let $p_k$ denote the position of the kink and $p_{k-1}$ and $p_{k+1}$ the respective positions of the antikinks. For simplicity we set $l_{k-1}=l_{k}$, and assume that $l_{k-2}, \; l_{k+1}>>l_k$. As in the case of kink binary interaction, Eqs.~(\ref{eom1}) and (\ref{eom2}) can be used to obtain dynamical equations for $\dot{p}_{k-1}$, $\dot{p}_{k}$, and $\dot{p}_{k+1}$. They are as follows.    
\begin{align}\label{ternaryk}
    \dot{p}_{k-1}=v_a e^{-2s_-l_k}, \;  \dot{p}_k=0, \; \dot{p}_{k+1}=-v_a e^{-2s_-l_k},
\end{align}
where $v_a$ is given by Eq.~(\ref{vkva}). 
We obtain equations of motion for the phase boundaries in a type-A ternary by proceeding in a similar fashion. Let $p_a$ be the position of the antikink and $p_{a-1}$ and $p_{a+1}$ the respective positions of the kinks. We set $l_{a-1}=l_a$ and assume that the ternary is isolated, i.e., $l_{a-2}, l_{a+1} >>l_a$. Using Eq.~(\ref{eom1}) and (\ref{eom2}) as in the previous case leads to   
\begin{align}\label{ternarya}
     \dot{p}_{a-1}=v_k e^{-2s_-l_a}, \;  \dot{p}_a=0, \; \dot{p}_{a+1}=-v_k e^{-2s_-l_a},   
\end{align}
where $v_k$ is given by Eq.~(\ref{vkva}). 

 The following comparisons of the two types of ternaries are in order here. Recall that $v_k$ increases and $v_a$ decreases as the driving strength $\e$ increases, and for large values of $\e$, $v_a \simeq 0$, and $v_k$ is large. This is evident from Fig.~\ref{fig:vkva}. It follows from Eqs.~(\ref{ternaryk}) and (\ref{ternarya}) that for large values of $\e$, the antikinks in type-K ternary are almost stationary, and the kinks in type-A ternary moves at a high speed. Therefore, in a system with multiple phase boundaries we expect type-A coalescence to dominate over type-K coalescence.  Here, type-A coalescence refers to the coalescence of a ternary of type-A resulting in a kink, and type-K coalescence refers to the coalescence of a ternary of type-K resulting in an antikink. We note here that this feature bear a resemblance with the cCH equation, where only type-A coalescence is allowed~\citep{Watson_2003}.   

\subsection{Simulation of kink binary coalescence.}
Here we discuss the specifics of the simulations and state the results. Kink binaries are numerically time evolved at different values of the driving strength~$\e$. For this a domain of length $L_D=50$ is considered such that $x\in(0, L_D).$ The initial profile is constructed using the $\tanh$ functions as shown in Eq.~(\ref{ansatz}) and is written as 
\begin{equation}
    \psi(x)=\Omega_1(x-p_1)+\Omega_2 (x-p_2) -1,
\end{equation}
where $\Omegai$ is given by Eq.~(\ref{CapitalOmega}). The variables $p_1$ and $p_2$ denote the initial positions of the kink and the antikink, respectively. The initial profile is numerically evolved in time according to the dAC equation with the help of Mathematica. The simulation is run for three different driving strengths, $\e=0.1$, $1$, and $1.5$. The results are shown in Fig.~\ref{fig:simu}, and the initial conditions for the different runs are stated therein.
\begin{figure}
    \centering
    \includegraphics[scale=0.7]{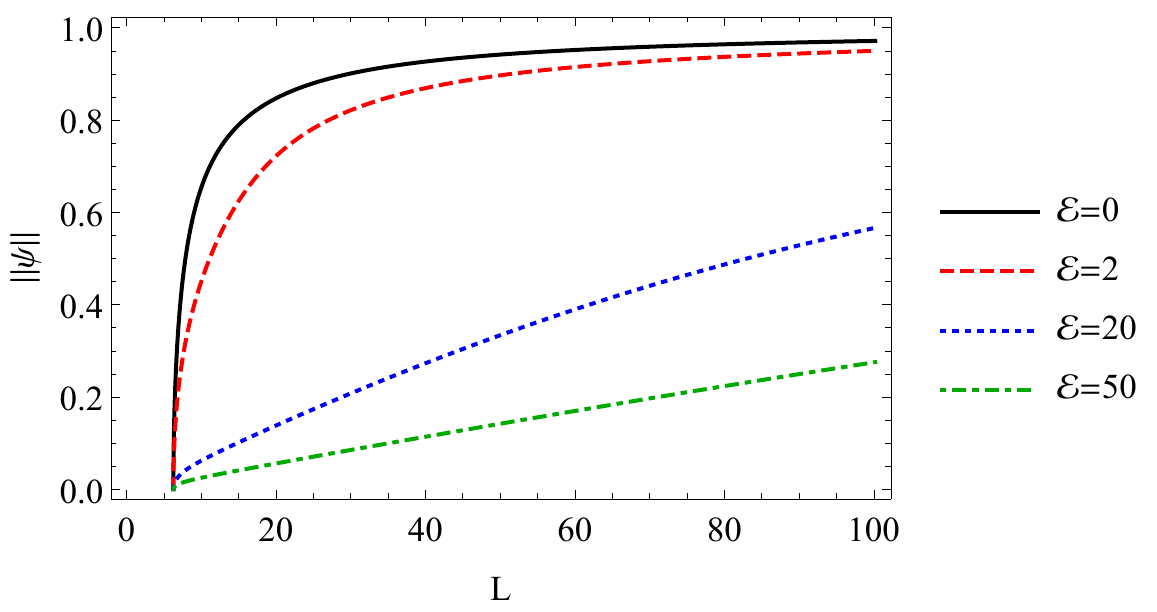}
    \caption{Branches of one-period stationary solutions of the dAC equation corresponding to different values of $\mathcal{E}$.}
    \label{fig:bifurcation}
\end{figure}
As can be seen from the figures, in all the three cases, the kink moves at a higher speed than the antikink. As $\e$ increases, the speed of the kink increases, whereas the speed of the antikink tends to $0$. For the cases of $\e=1$ and $\e=1.5$, there is no visible shift in the position of the antikink.  Further, the time taken for coalescence falls from $15537$ to $1457$ when $\e$ is increased from $1$ to $1.5$. Note that in both cases the initial separations between the kink and the antikink are the same: $l_1=p_2 -p _1 =15$. 
 
 \section{One and two period stationary solutions and coarsening modes}\label{sec:oneperiod}

 In the first part of this section, we focus on the stationary solutions of the dAC equation~(\ref{cAC}) subject to periodic boundary conditions. In particular, we seek the solutions of the equation
 \begin{equation}\label{cAC0}
    \psi_{xx}+\psi- \psi^3 + 2\e \psi \psi_x=0,
\end{equation}
 on a periodic domain of size $L$ such that  $x\in(0,\,L).$
 We limit to the case of one-period solutions with 0 mean for simplicity. The solutions are obtained by numerical continuation, implemented with the help of the continuation and bifurcation software Auto07p~\citep{Auto07p}. We also draw the bifurcation diagrams and study the stability of the one-period solutions. In the later part of this section, we focus on the linear stability of the two-period solutions. We therein obtain the positive eigenvalues and the corresponding eigenfunctions (coarsening modes) and study their behavior as a function of the domain size $L$ and the driving strength $\e$.  
\begin{figure}
    \centering
    \includegraphics[scale=0.7]{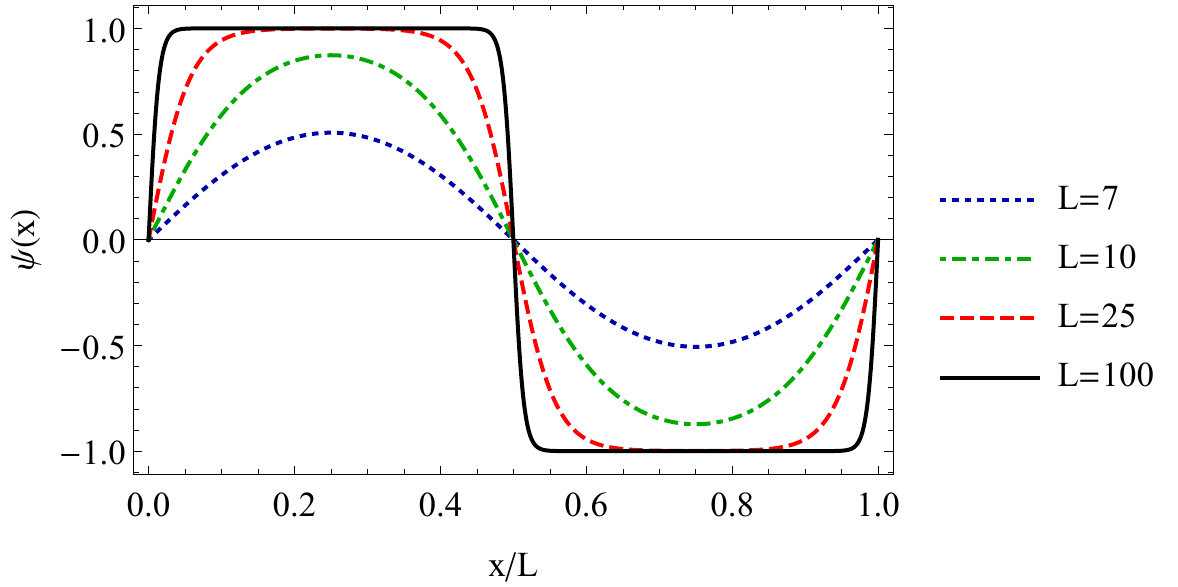}
    \caption{One-period stationary solutions of the dAC equation corresponding different values of $L$ for the case of $\mathcal{E}=0$.}
    \label{fig:stdacsoln}
\end{figure}
\subsection{One-period solutions}
We rewrite Eq.~(\ref{cAC0}) as a set of two first order differential equations in terms of the scaled coordinate $z=x/L$ for convenience.  
\begin{align}\label{cACcoupled}
    \psi_z&=L \widetilde{\psi}, \nonumber \\
    \widetilde{\psi}_z & = -L \left( \psi -\psi^3 + 2 \e \psi \widetilde{\psi} \right), 
\end{align}
where $\widetilde{\psi}\equiv \psi_x$ and  $z\in(0,\,1)$. Note that in Eq.~(\ref{cACcoupled}) we have extracted the period $L$ as a parameter.
We first make the following preliminary observations in order to implement numerical continuation and obtain the one-period solutions. The spatially uniform solutions $\psi_0=\pm 1$ are stable with respect to time dependent perturbations for any domain size $L$ as shown in Sec.~\ref{asymptotic}. However, in the case of the constant solution $\psi_0=0$, modes with $|k|<k_c=1$ are unstable while those with $|k|>k_c$ are stable, where $k$ is the wave vector [see Eq.~(\ref{betareal})]. This means that the unstable modes step in when the length of the domain $L$ is increased beyond $L_c=2 \pi/k_c=2 \pi$. We identify the point $L=L_c$ as the primary bifurcation point. 

We now set the initial value of the parameter $L=L_c$ and choose the following small amplitude sinusoidal function as the starting solution.
\begin{align}
    \psi\left(x(z)\right)&=0.0001 \sin(2 \pi z), \nonumber \\
    \widetilde{\psi}\left(x(z)\right)&=0.0001 \cos(2 \pi z). 
\end{align}
The branches of one-period solutions corresponding to different values of the driving strength $\e$ are now obtained from this initial data by varying $L$ as the continuation parameter. This is implemented using the software Auto07p~\citep{Auto07p}. Note that the boundary conditions
\begin{align}
    \psi\left(x(0))\right)=\psi\left(x(1)\right), \nonumber \\
    \widetilde{\psi}\left(x(0))\right)=\widetilde{\psi}\left(x(1)\right),
\end{align}
\begin{figure}
    \centering
    \includegraphics[scale=0.7]{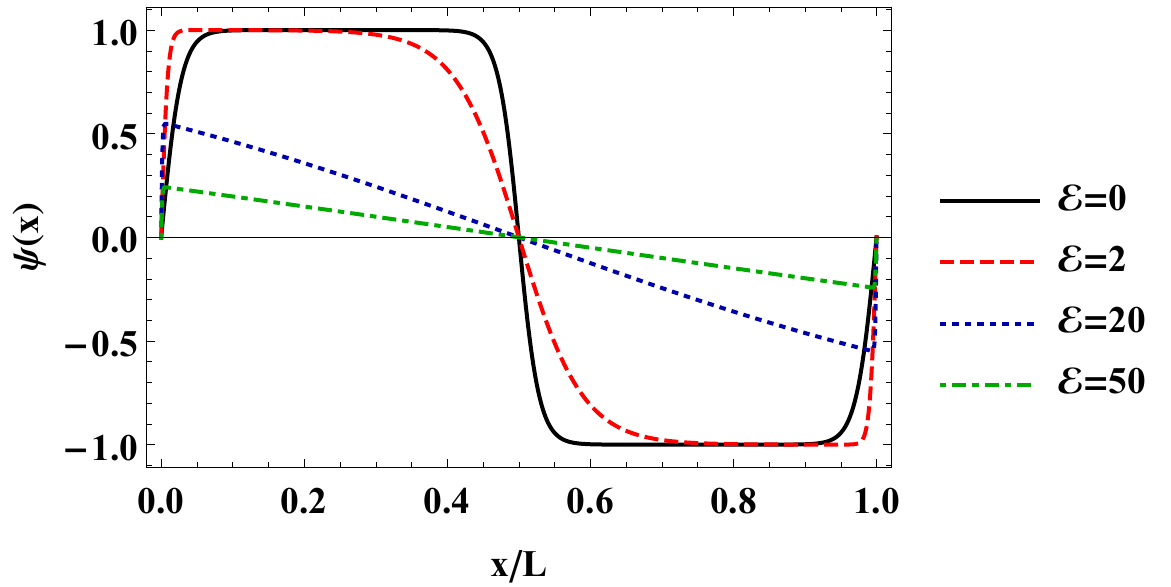}
    \caption{Stationary one-period solutions of the dAC equation corresponding to different values of the driving strength $\e$ for $L=50$.}
    \label{fig:psiofx_pbc}
\end{figure}
and the additional constraint 
$$\int_0^1 \! \!  dz \; \psi =0,$$ 
are also imposed. For more details on implementation of such calculations and examples see Ref.~\citep{Tseluiko_2020,Thiele_autoeg, thiele2014munsteranian}.

 For each value of $\e$ a branch of spatially non-uniform solutions emerges from the primary bifurcation point $L_c$ as shown in Fig.~\ref{fig:bifurcation}. These solutions are characterized using their norms
\begin{equation}
    ||\psi||= \sqrt{\frac{1}{L}\int_0^L \! \! \! dx \;\psi^2}.
\end{equation}
We show using the weakly non-linear analysis presented in Appendix~\ref{appendix:landau} that the primary bifurcation at $L=L_c$ is super critical for all values of $\mathcal{E}$. Moreover, we find no further bifurcations along the solution branches shown in Fig.~\ref{fig:bifurcation}.

The case of $\mathcal{E}=0$ corresponds to that of the standard Allen-Cahn equation. The solutions at five different points on the $\e=0$ branch are plotted in Fig.~\ref{fig:stdacsoln}. The solution profiles resemble a sinusoidal wave  for values of $L$ close to $L_c$. But as $L$ increases plateaus where $\psi=+1 \text{ and }-1$ appear separated by a narrow anti-kink. 

We now fix the period at $L=50$ and continue along the parameter $\mathcal{E}$. The solutions corresponding to four different values of $\e$ are plotted in Fig.~\ref{fig:psiofx_pbc}. It evident from the figure that when the driving strength $\mathcal{E}$ is switched on the anti-kink region begins to widen and the plateaus become narrower. At the same time the kink at the boundaries become sharper. Eventually, the plateaus vanish  and the anti-kink region becomes straight line when $\e$ is sufficiently large.  When $\mathcal{E}$ is further increased the profile tends to flatten out. This is in contrast with the case of the cCH equation, where the plateaus assume a spatially irregular profile when the driving force is increased beyond a certain value, and there are also other periodic stationary solutions when the drive is large~\citep{Golovin_1998, Tseluiko_2020}. 
\begin{figure}
    \centering
    \includegraphics[scale=0.53]{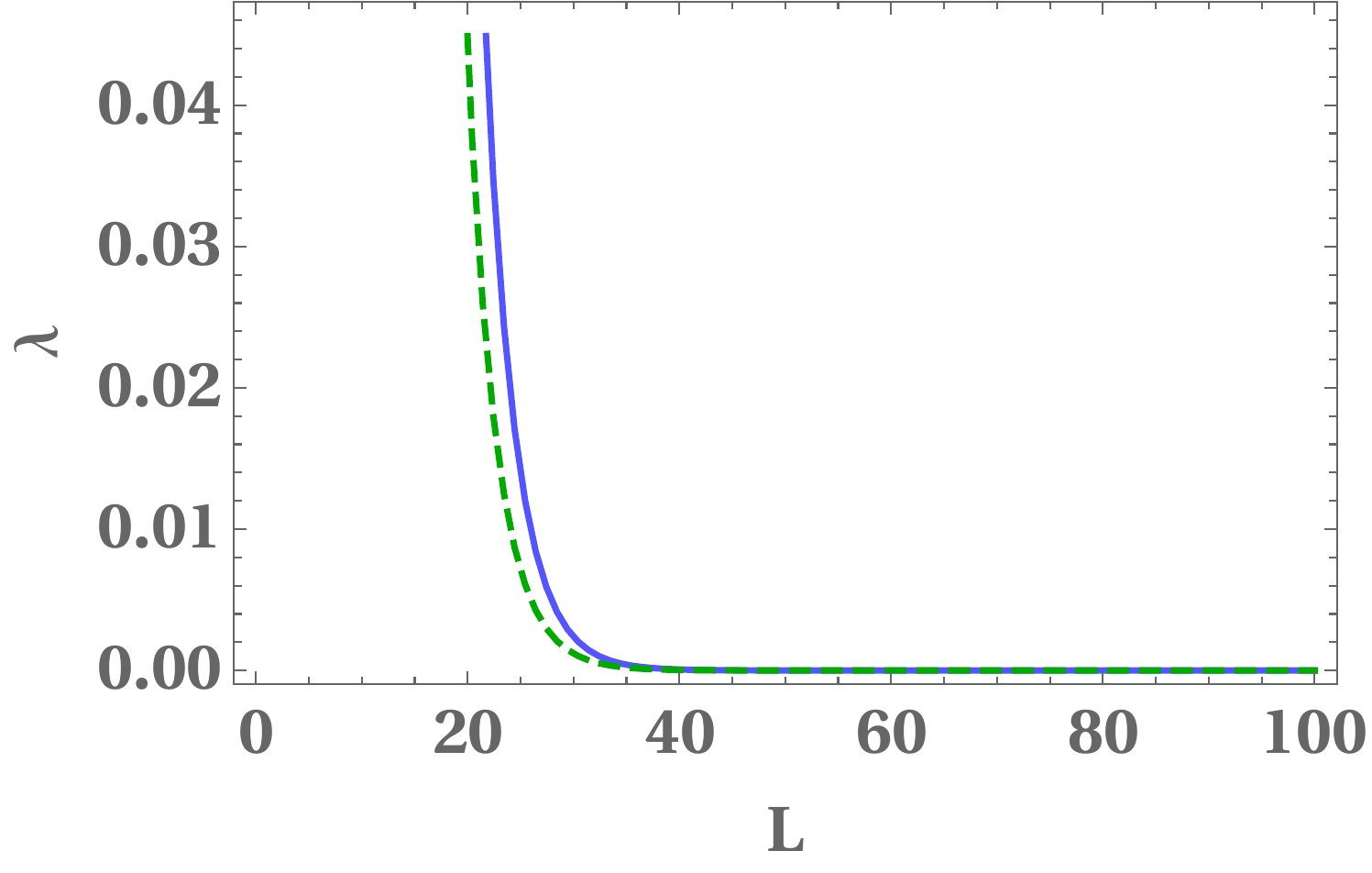}
    \caption{The dominant (blue solid line) and the non-dominant (green dashed line) eigenvalues against $L$ for $\e=0.$}
    \label{fig:evalvsl}
\end{figure}
\subsection{Two-period solutions and coarsening modes.}
\begin{figure*}
\subfloat[$\e=0$, $L=30$. Mode type: $M_B$ \label{subfig:2pe0dom}]{
\includegraphics[scale=0.363]{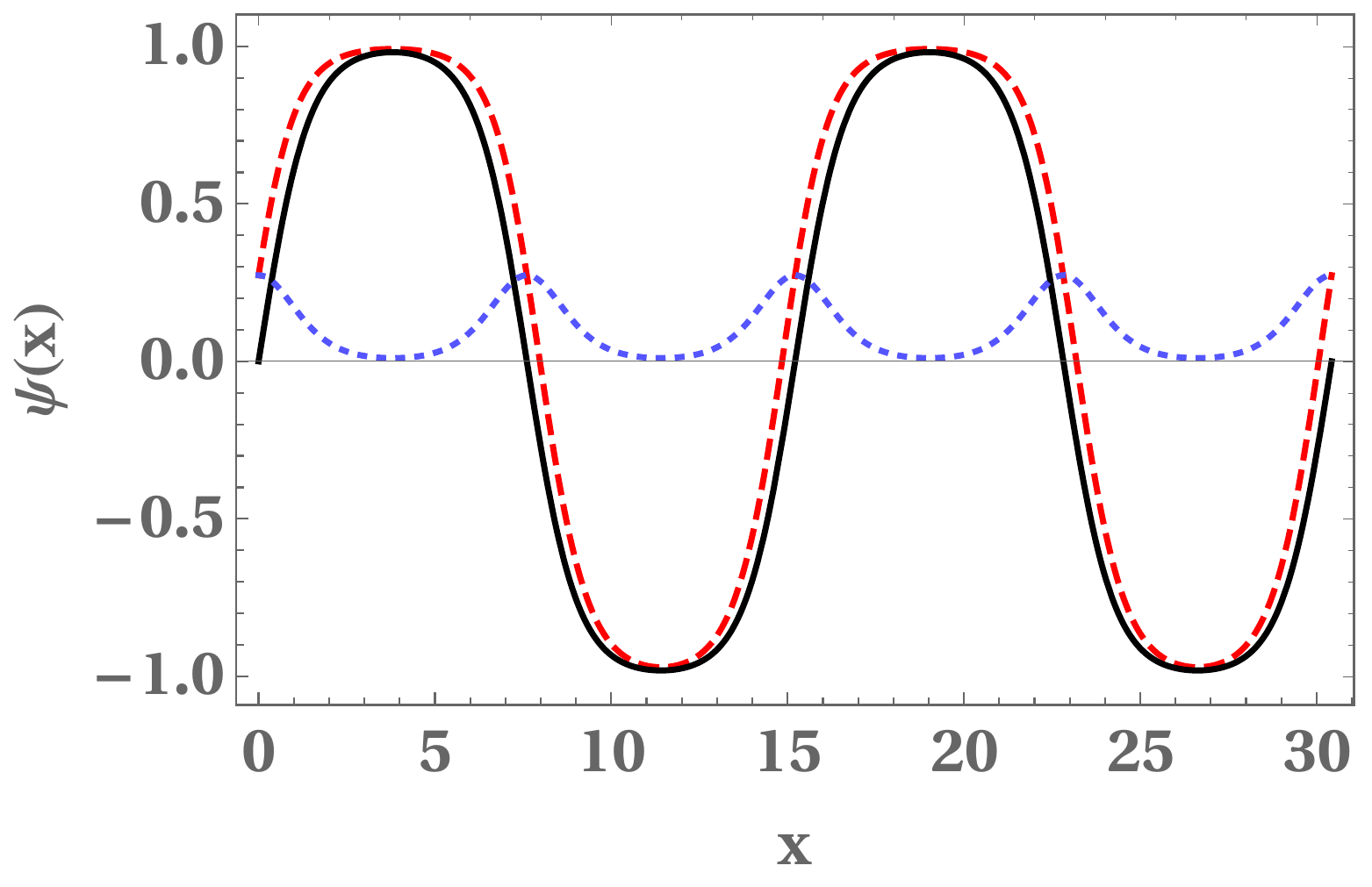}} 
\subfloat[$\e=0$, $L=30$. Mode type: $M_A$ \label{subfig:2pe0subdom1}]{
\includegraphics[scale=0.363]{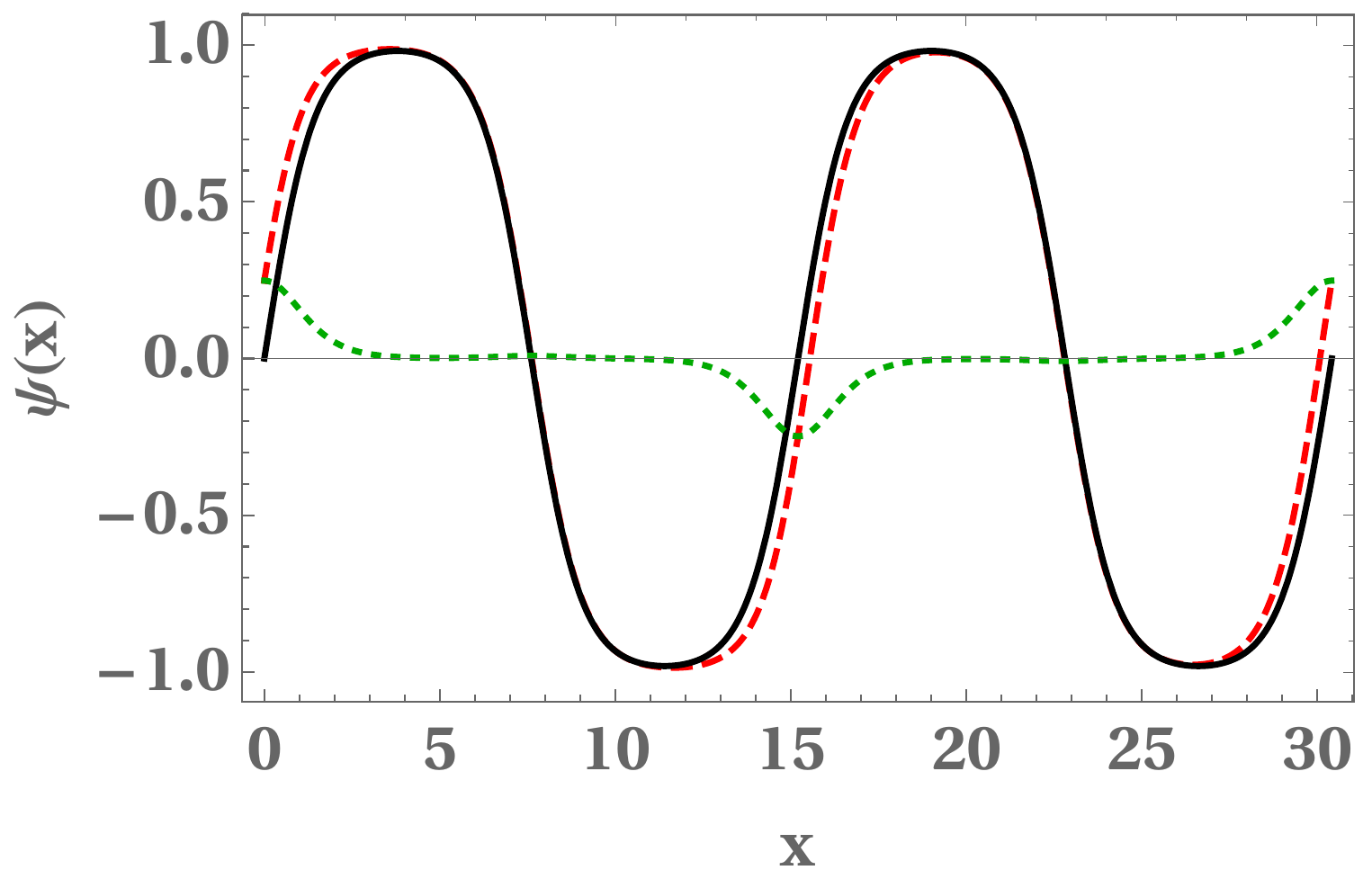}}
\subfloat[$\e=0$, $L=30$. Mode type: $M_K$ \label{subfig:2pe0sumdom2}]{
\includegraphics[scale=0.363]{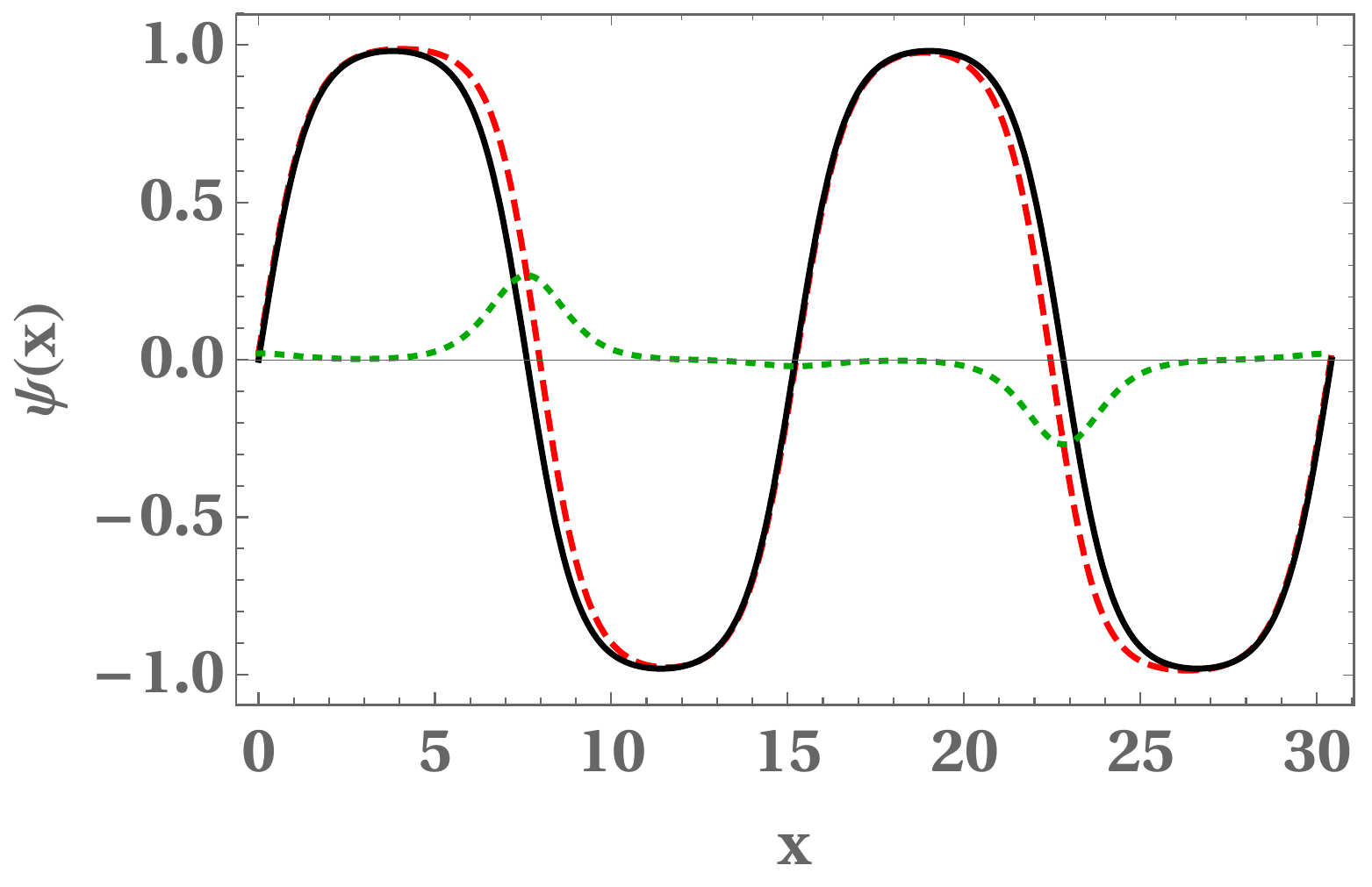}}\\
\subfloat[$\e=0.08$, $L=30$. Mode type: $M_B$\label{subfing:2pe0p16}]{
\includegraphics[scale=0.45]{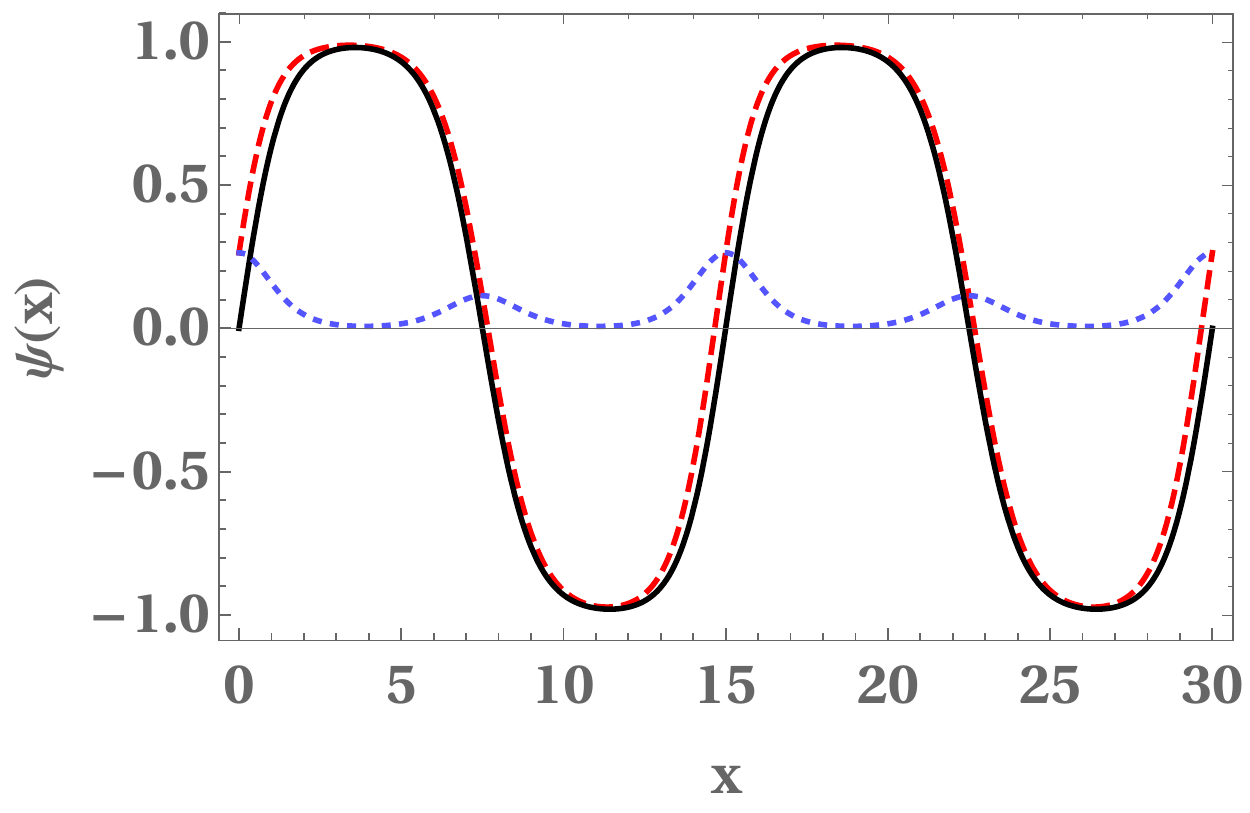}}
\subfloat[$\e=0.75$, $L=30$. Mode type: $M_B$\label{subfig:2pe1p5}]{
\includegraphics[scale=0.45]{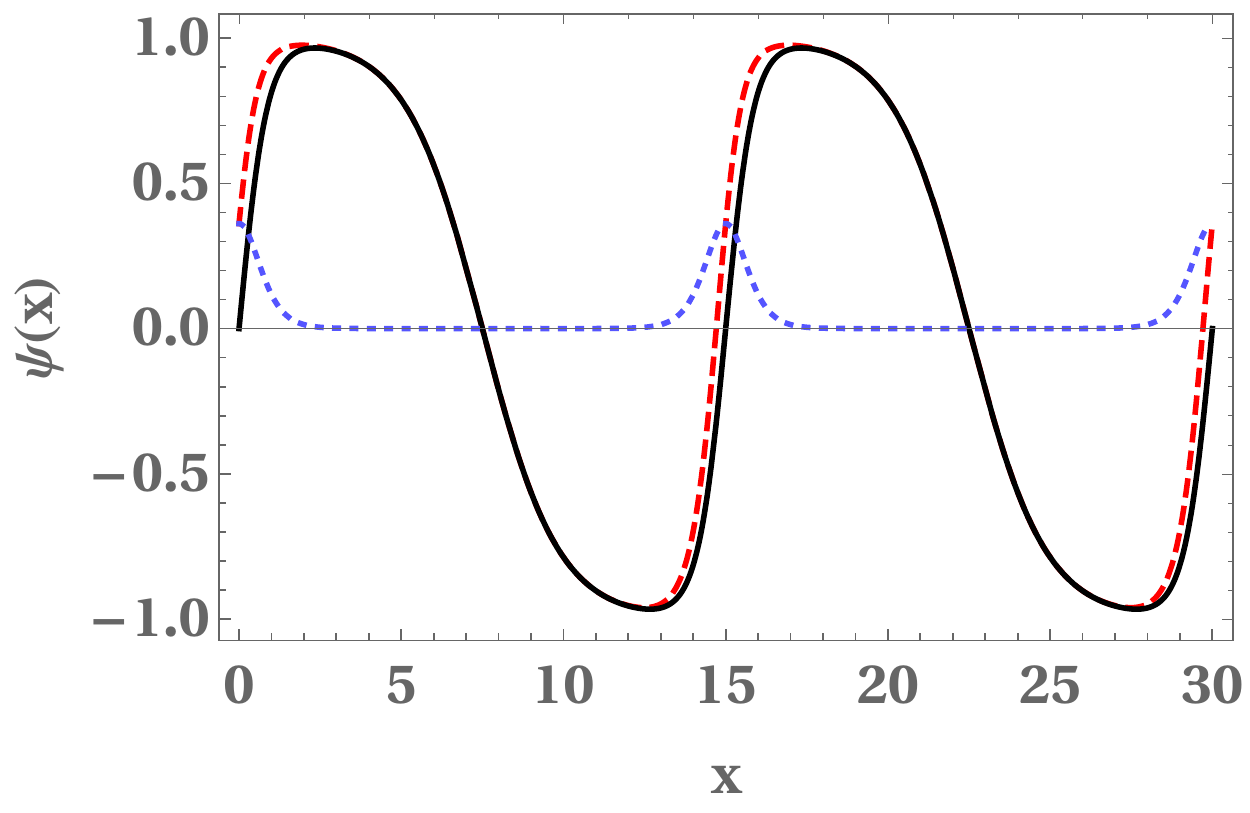}}
\subfloat[Positive eigenvalues against $\e$, for $L=30$\label{subfig:evalsvse}]{
\includegraphics[scale=0.45]{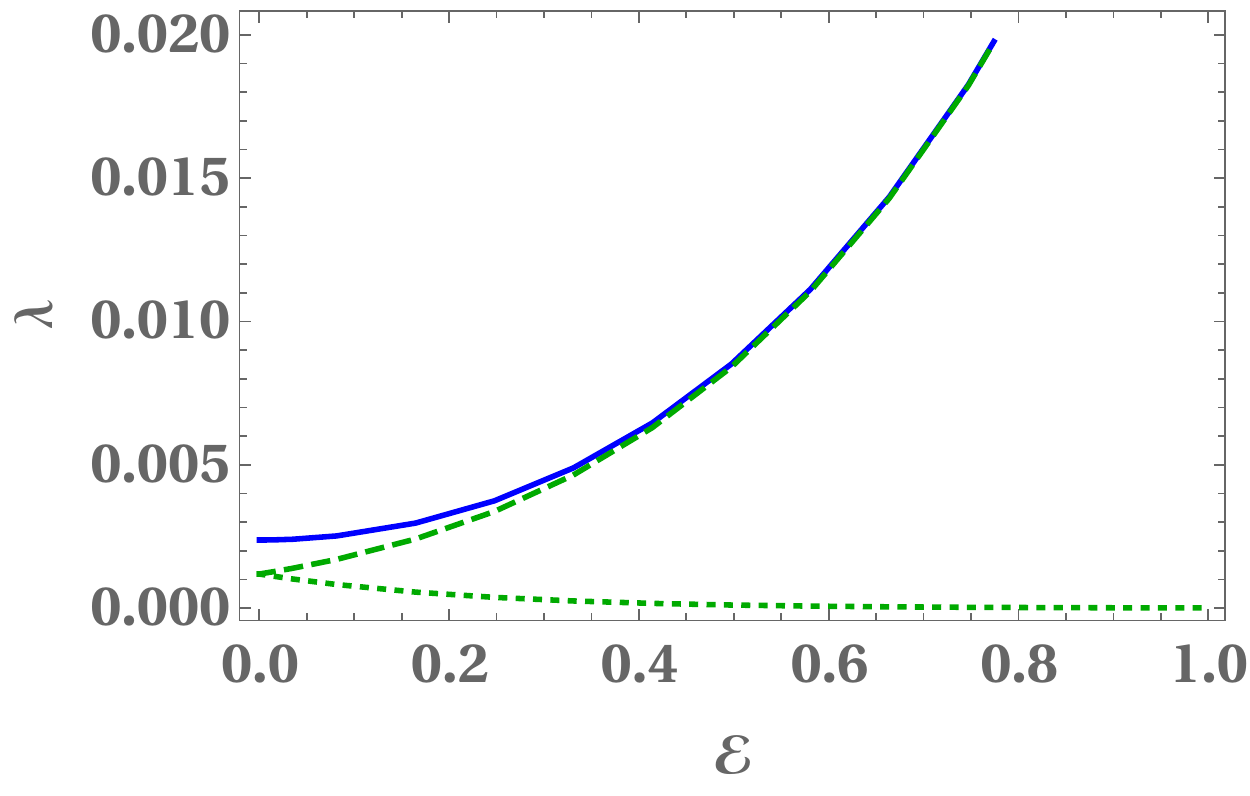}}
    \caption{Each of the first five figures a) to e) shows the two period solution $\psi_s$ (black solid line), the coarsening mode $\delta \psi$ (blue/green dotted line), and the superposition $\psi_s+\delta\psi$ (red dashed line) corresponding to the type of mode and values of $L$ and $\e$ given under them. The last figure f) shows the behavior of the $M_B$, $M_K$, and $M_A$ eigenvalues as a function of $\e$ in blue solid, green dashed, and green dotted lines, respectively. }
    \label{fig:coarsening}
\end{figure*}

 The two period solutions can be obtained by numerical continuation or constructed from the one-period solutions obtained above. For instance, a two-period solution corresponding to the parameter values $(L, \e)$ can be constructed by concatenating two identical one-period solutions with the parameter values $(L/2,\e)$. See the Figs.~\ref{subfig:2pe0dom}, \ref{subfing:2pe0p16}, and \ref{subfig:2pe1p5} for the two-period solutions corresponding to $L=30$ and $\e=0, \; 0.08, \; \text{and} \; 0.75$, respectively. We will find from the linear stability analysis described in the next paragraph that the two period solutions are unstable to coarsening modes. 

Let $\psi_s(x)$ be a two-period stationary solution of Eq.~(\ref{cAC}) and $\delta \psi(x,t)$ a perturbation. Substituting $\psi=\psi_s+\delta \psi$ in Eq.~(\ref{cAC}) and collecting terms that are linear in $\delta \psi$ yields 
\begin{equation}
   \partial_t \delta \psi=\hat{\mathcal{L}} \delta \psi,
\end{equation}
where the operator
\begin{align}
 \hat{\mathcal{L}} &= \sum_{k=0}^2 \mathcal{F}_k(x) \partial_x ^k, \nonumber \\
     \mathcal{F}_0 & = (1-3\psi_s^2)+2\e \partial_x \psi_s, \nonumber \\
     \mathcal{F}_1 & =v+2 \e \psi_s,
    \nonumber \\
    \mathcal{F}_2&=1.
\end{align}
The stability of the solution $\psi_s$ is determined by the  eigenvalues of operator $\hat{\mathcal{L}}$, which is a linear ordinary differential operator with non-constant coefficients. The eigenfunctions and the corresponding eigenvalues can be computed numerically using several methods. We use the Floquet-Fourier-Hill method~\citep{Floquet_fourier}. We now proceed to examine in detail the coarsening modes and the corresponding eigenvalues thus obtained.

Consider the case of $\e=0$. Recall that Eq.~(\ref{cAC}) reduces to the standard Allen-Cahn equation when $\e=0$. In this case there are two positive eigenvalues, which we denote by $\lambda.$ They are plotted against $L$ in Fig~\ref{fig:evalvsl}. We call the larger of these the dominant eigenvalue and the other the non-dominant eigenvalue. However, the difference between these eigenvalues decreases as $L$ increases and they both asymptotically approach $0$.

The dominant eigenvalue is non-degenerate while the non-dominant eigenvalue is doubly degenerate. 
The eigenfunctions corresponding to these eigenvalues, i.e., the positive eigenvalues, are referred to as the coarsening modes. We illustrate the case of $L=30$ in Fig.~\ref{fig:coarsening} to describe the nature of these modes. In particular, the blue dotted line in Fig.~\ref{subfig:2pe0dom} shows the dominant coarsening mode, i.e., the eigenfunction corresponding to the dominant eigenvalue. The two non-dominant coarsening modes corresponding to the non-dominant eigenvalue are plotted using green dotted lines in Fig.~\ref{subfig:2pe0subdom1} and \ref{subfig:2pe0sumdom2}, respectively. The red dashed line in each of these figures is used to illustrate the superposition
\begin{equation}
    \psi=\psi_s + \epsilon \delta \psi,
\end{equation}
where $\delta \psi$ represents  the respective coarsening mode, $\psi_s$ is the two-period solution and $\epsilon$ is a small number. As is now evident from the figure, the dominant coarsening mode correspond to kink binary interaction where adjacent kinks and antikinks move towards each other and annihilate. We refer to this type of modes as $M_B$. The first of the non-dominant coarsening mode (shown in Fig.~\ref{subfig:2pe0subdom1}) corresponds to the process where kinks on either side of an antikink move towards it and coalesce resulting in a kink, and the second non-dominant coarsening mode (shown in Fig.~\ref{subfig:2pe0sumdom2}) corresponds to the process where antikinks on either side of a kink move towards it and coalesce resulting in an antikink. We refer to modes of the former type as $M_A$ and the latter type as $M_K$.   

We now examine the behavior of the positive eigenvalues and the coarsening modes when $\e>0.$ The main observations are listed below. 
\begin{enumerate}
\item There are three types of coarsening modes as in the case of $\e=0$: $M_B$, $M_A$, and $M_K$. 

\item $M_A$-type and $M_K$-type modes become non-degenerate as soon as the driving strength~$\e$ is switched on. The $M_A$ and the $M_K$ eigenvalues are plotted against~$\e$ in Fig.~\ref{subfig:evalsvse} using green dashed and green dotted lines, respectively. As shown therein, the $M_A$ eigenvalue increases with $\e$ and asymptotically becomes equal to the $M_B$ eigenvalue, whereas the $M_K$ eigenvalue decreases with $\e$ and asymptotically tends to 0. 

\item $M_B$-type mode remains the only dominant coarsening mode except when $\e$ is large.
This is evident from Fig.~\ref{subfig:evalsvse} where the $M_B$ eigenvalue is plotted against the driving strength $\e$ using the blue solid line. Both $M_B$ and $M_A$-type modes are dominant when $\e$ is large as the corresponding eigenvalues tend to be equal then.  

\item The behavior of the $M_B$-type mode varies with $\e$: as $\e$ increases the antikinks tend to be stationary. This is evident on comparing the Figs.~\ref{subfig:2pe0dom}, \ref{subfing:2pe0p16}, and \ref{subfig:2pe1p5} which correspond to the cases of $\e=0$, $\e=0.08$, and $\e=0.75$, respectively. The domain size $L=30$ for all the three cases. The blue dotted line in each of the three figures shows the corresponding $M_B$-type coarsening mode, and the red dashed line the superposition $\psi=\psi_s+\epsilon \delta\psi.$ 
\item The $M_K$-type coarsening mode diminishes as $
\e$ increases, and $\delta \psi \simeq 0$ for large $\e$.  The investigation here imply that type-K coalescence is favoured over type-A at large $\e$, which is in agreement with the inferences of Sec.~\ref{sec:nneighbour}.  Remember that type-K (type-A) coalescence refers to the one where two kinks (antikinks) meet an antikink (kink) resulting in a kink (antikink).  
\end{enumerate}

To summarize, in the first part of this paper, we studied the single (anti)kink steady state solutions of the dAC equation in an infinite domain and established that such solutions exist only for isolated values of the traveling wave velocity $v.$ We also investigated therein the linear stability of the constant solutions. We then considered a driven Allen-Cahn system with multiple phase boundaries and derived equations that govern the dynamics of these boundaries when the driving strength $\e$ and the separation between the phase boundaries $l_i$ are large. Using these equations, we investigated various coarsening mechanisms and studied their behavior as a function of the driving strength $\e.$ We further argued that the average domain size scaled logarithmically with respect to time. We also presented the results of the simulations of kink binary coalescence and compared them with the analytical results. The time taken for kink binary coalescence is found to decrease drastically with increase in~$\e.$ 

In the last section, we presented the bifurcation analysis of the one-period stationary solutions of the dAC equation and explored their behavior with respect to the period $L$ and the driving strength $\e.$ We then investigated the linear stability of the two-period solutions and thereby identified the various coarsening modes. We observed, among other details, that the $M_B$-type mode is the only dominant coarsening mode for small $\e$ and that for large $\e$ the $M_K$ eigenvalue asymptotically becomes equal to the $M_B$ eigenvalue. 

\section{Acknowledgment}

I thank Sreedhar B. Dutta for helping me in formulating the problem and engaging in fruitful discussions during the progress of the work and the preparation of the manuscript. 

\appendix
\begin{widetext}
\section{Approximating the integrals in Eq.~(\ref{nnnoapprox})}\label{appendix:integrals}
In this appendix, we evaluate the integrals appearing on the left hand side (LHS) and the right hand side (RHS) of Eq.~(\ref{nnnoapprox}) using the approximations explained below. 
\begin{enumerate}
    \item It follows from Eq.~(\ref{sis}) that when the driving strength $\e$ is large $s_+$ is also large. Then, the function $\sech^2[s_+(x-p_k)]$ becomes sharply peaked at $p_k$ and approximately $0$ elsewhere. This fact is exploited in approximating the integrals in which $\sech^2[s_+(x-p_k)]$ appears. For instance, the integrals of the form
    \begin{equation}\label{form}
      I=  \int_{-\infty}^{\infty} \! \! \! dx \; \sech^2[s_+(x-p_k)] \, g\boldsymbol{(}s_+(x-p_k)\boldsymbol{)} \, f(x),
    \end{equation}
   are approximated as
    \begin{equation}\label{approx}
               I\simeq \int_{-\infty}^{\infty} \! \! \! dx \; \sech^2[s_+(x-p_k)] \, g\boldsymbol{(}s_+(x-p_k)\boldsymbol{)} \, \left\{f(p_k)+f'(p_k) (x-p_k) \right\},
    \end{equation}
    where $f(x)$ is Taylor expanded about $p_k$ to first order in $(x-p_k).$
    
    \item Let $p_i$ and $p_j$, where $j=i\pm1$, denote positions of two adjacent phase boundaries. The following approximations can be made when $|p_i-p_j|$ is large: 
    \begin{align}\label{approx2}
        \tanh[s_\pm (p_i-p_j)] &\simeq 
        \begin{cases}
           -1+2e^{-2 s_\pm l_i}, & j=i+1 \\
            1-2e^{-2 s_\pm l_j}, & j=i-1,
        \end{cases} \nonumber \\
        \sech^2[s_\pm (p_i-p_j)] &\simeq 
        \begin{cases}
           4e^{-2 s_\pm l_i}, & j=i+1 \\
           4e^{-2 s_\pm l_j}, & j=i-1,
        \end{cases}
    \end{align}
    where $l_i$ is given by Eq.~(\ref{li}) and  terms of $O(\{\exp(-2s_\pm l_i)\}^2)$ and higher order are neglected. 
\end{enumerate}

We now proceed to approximate the integrals in Eq.~(\ref{nnnoapprox}). Using Eq.~(\ref{CapitalOmega}) the LHS can be explicitly written as
\begin{align}\label{step1}
    \text{LHS}=- \sum_{j=-1}^1   \!\dot{p}_{i+j}\int_{-\infty}^{\infty} \! \! dx \, \Omega^i_x \Omega^{i+j}_{x}  = (-1)^{j+1} s_i \sum_{j=-1}^1 s_{i+j} \dot{p}_{i+j}\int_{-\infty}^{\infty} \! \! dx \, \sech^2{(s_i \zeta_i)} \, \sech^2(s_{i+j} \zeta_{i+j}).
\end{align}
We consider the cases of odd and even $i$ separately. Note that in the former case, $i$ represents a kink, and in the latter case, an antikink, by the convention adopted in Sec.~\ref{sec:nneighbour}. When $i$ is odd, we can write Eq.~(\ref{step1}) as
\begin{align}\label{lhsoddraw}
    \text{LHS (odd)}=-s_+^2 \dot{p}_{i} \int_{-\infty}^{\infty} \! \! dx \, \sech^4(s_+ \zeta_{i}) \;
    +s_+ s_- \sum_{j \in \{-1,1\}} \dot{p}_{i+j} \int_{-\infty}^{\infty} \! \! dx \, \sech^2(s_+ \zeta_{i}) \sech^2(s_- \zeta_{i+j}).
\end{align}
The integral in the first term is readily solved after replacing $\zeta_i$ using Eq.~(\ref{zeta}) to yield 
\begin{equation}\label{lhsodd1}
    \int_{-\infty}^{\infty} \! \! dx \, \sech^4 (s_+ \zeta_i)=\int_{-\infty}^{\infty} \! \! dx \, \sech^4 [s_+ (x-p_i)]=-\frac{4}{3}s_+ \dot{p}_i.
\end{equation}
The integral in the second term is of the form given in Eq.~(\ref{form}). Therefore this integral can be approximated in the same way, i.e., by Taylor expanding $\sech^2[s_-(x-p_{i+j})]$ about $p_i$ up to terms linear in $(x-p_i)$.
\begin{align}\label{lhsodd2}
    \int_{-\infty}^{\infty} \! \! dx \, \sech^2(s_+ \zeta_{i}) \sech^2(s_- \zeta_{i+j}) & \simeq 
        \int_{-\infty}^{\infty} \! \! dx \, \sech^2 [s_+ (x-p_i)] \left\{\sech^2[s_-(p_i-p_{i+j})] + (x-p_i) \frac{\partial}{\partial x} \sech^2[s_-(x-p_{i+j})]\right\} \nonumber \\
        &=\frac{2}{s_+}\sech^2[s_-(p_i-p_{i+j})] \nonumber \\
        &\simeq \frac{8}{s_+} e^{-2js_-(p_{i+j}-p_i)}.
\end{align}
Note that we have used Eq.~(\ref{approx2}) in the last step. Plugging the results from Eqs.~(\ref{lhsodd1}) and (\ref{lhsodd2}) in Eq.~(\ref{lhsoddraw}) and then using Eq.~(\ref{li}), we obtain 
\begin{equation} \label{lhsodd}
    \text{LHS (odd)} \simeq -\frac{4}{3}s_+ \dot{p}_i
    + 8 s_- \dot{p}_{i+1} e^{-2 s_- l_i} 
     + 8 s_- \dot{p}_{i-1} e^{-2 s_- l_{i-1}} 
\end{equation}
Proceeding in similar fashion, we arrive at the following approximation for the LHS when $i$ is even.
\begin{equation}\label{lhseven}
    \text{LHS (even)} \simeq -\frac{4}{3}s_- \dot{p}_i+8 s_- \dot{p}_{i+1}e^{-2 s_-l_i}+8 s_- \dot{p}_{i-1}e^{-2 s_-l_{i-1}}.
\end{equation}
The integrals in the RHS of Eq.~(\ref{nnnoapprox}) can be evaluated using the same method. Using Eqs.~(\ref{FF}) and (\ref{Ftilde}) the RHS is expanded as
\begin{align}\label{rhsraw}
 \text{RHS}=&\sum_{\alpha=\pm} \left[
 -3 \int_{-\infty}^{\infty} \! \! dx \, \Omegaid \Omegai \omega_\alpha^2 -\int_{-\infty}^{\infty} \! \! dx \, \Omegaid \omega_\alpha^3 +2 \e \int_{-\infty}^{\infty} \! \! dx \, \Omegaid \omega_\alpha ({\omega_{\alpha}}_x +\Omegaid) + 4 \e \int_{-\infty}^{\infty} \! \! dx \, \Omegaid \Omegai {\omega_\alpha}_x \right] \nonumber \\
 &-6 \int_{-\infty}^{\infty} \! \! dx \, \Omegaid \Omegai \omega_- \omega_+
 -3 \int_{-\infty}^{\infty} \! \! dx \, \Omegaid \omega_-^2 \omega_+ -3 \int_{-\infty}^{\infty} \! \! dx \, \Omegaid \omega_- \omega_+^2 +2 \e \int_{-\infty}^{\infty} \! \! dx \, \Omegaid ( \omega_+ {\omega_-}_x + \omega_- {\omega_+}_x).
\end{align}
 We consider the cases of odd and even $i$ separately as in the case of LHS. The first integral in the right hand side of Eq.~(\ref{rhsraw}) can be explicitly written for odd $i$ as
\begin{align}\label{stepinta1}
I_1 (\text{odd})=\int_{-\infty}^{\infty} \! \! dx \, \Omegaid \Omegai \omega_\alpha^2=
s_+ \int_{-\infty}^{\infty} \! \! dx \, \sech^2[s_+(x-p_i)] \tanh[s_+ (x-p_i)] \left\{\tanh[s_-(x-p_{i+a_\alpha})]+a_\alpha\right\}^2,
\end{align}
where we have used Eqs.~(\ref{sis}), (\ref{zeta}), (\ref{CapitalOmega}),  and (\ref{smallomega}) and $a_\pm=\pm1$. The integral in Eq.~(\ref{stepinta1}) is of the form given in Eq.~(\ref{form}) and can be approximated as shown in Eq.~(\ref{approx}). 
\begin{align}\label{oddi1}
 I_1(\text{odd})\simeq &s_+\int_{-\infty}^{\infty} \! \! dx \, \sech^2[s_+(x-p_i)] \tanh[s_+ (x-p_i)]
 \left[\left\{\tanh[s_-(p_i-p_{i+a_\alpha})]+a_\alpha\right\}^2 + \right. \nonumber \\
 &\left. 2s_-\left\{\tanh[s_-(p_i-p_{i+a_\alpha})]+a_\alpha\right\}\sech^2[s_- (p_i-p_{i+a_\alpha})] \times (x-p_i) \right] \nonumber \\
 =&2s_- s_+\left\{\tanh[s_-(p_i-p_{i+a_\alpha})]+a_\alpha\right\}\sech^2[s_- (p_i-p_{i+a_\alpha})]\int_{-\infty}^{\infty} \! \! dx \, \sech^2[s_+(x-p_i)] \tanh[s_+ (x-p_i)] (x-p_i) \nonumber \\
 =&\frac{2 s_-}{ s_+}\left\{\tanh[s_-(p_i-p_{i+a_\alpha})]+a_\alpha\right\}\sech^2[s_- (p_i-p_{i+a_\alpha})] \nonumber \\
 \simeq& \frac{8 s_-}{s_+} \times
 \begin{cases}
\left(e^{-2 s_-l_i} \right)^2, & \alpha=+ \\
    -\left(e^{-2 s_-l_{i-1}} \right)^2, & \alpha=-
 \end{cases} \nonumber \\
 \simeq &0. 
\end{align}
Note that in the one to last step we have used Eq.~(\ref{approx2}), and in the last step neglected terms of order $O(\left\{\exp{(-2s_{\pm}l_i)}\right\}^2)$. The remaining integrals in Eq.~(\ref{rhsraw}) are evaluated likewise for the case of odd $i$. The non-vanishing contributions are listed below. 

\begin{align}\label{oddi3}
  \int_{-\infty}^{\infty} \! \! dx \, \Omegaid \omega_\alpha ({\omega_{\alpha}}_x +\Omegaid)\simeq
  - \frac{8}{3}s_+\times
  \begin{cases}
    e^{-2 s_- l_i}, & \alpha =+ \\
    -e^{-2 s_- l_{i-1}}, & \alpha =-,
  \end{cases}
\end{align}
\begin{align}\label{oddi5}
 \int_{-\infty}^{\infty} \! \! dx \, \Omegaid \Omegai {\omega_\alpha}_x \simeq
    -8 s_- \frac{s_-}{s_+}\times
 \begin{cases}
    e^{-2 s_- l_i}, & \alpha =+ \\
    -e^{-2 s_- l_{i-1}}, & \alpha =-.
 \end{cases}
 \end{align}
Using the results from Eqs.~(\ref{oddi1}), (\ref{oddi3}), and (\ref{oddi5}) in Eq.~(\ref{rhsraw}) we obtain
\begin{equation}\label{rhsodd}
    \text{RHS (odd)} \simeq 16 \e \left( \frac{s_+}{3}+2 s_- \frac{s_-}{s_+} \right)e^{-2 s_- l_{i-1}}
    -16 \e \left( \frac{s_+}{3}+2 s_- \frac{s_-}{s_+} \right)e^{-2 s_- l_{i}}.
\end{equation}
We now proceed to evaluate the integrals in Eq.~(\ref{rhsraw}) for the case of even $i$. Using Eqs.~(\ref{sis}), (\ref{zeta}), (\ref{CapitalOmega}), and (\ref{smallomega}), the first integral is explicitly written for even $i$ as
\begin{align}
I_1 (\text{even}) =\int_{-\infty}^{\infty} \! \! dx \, \Omegaid \Omegai \omega_\alpha^2= 
s_- \int_{-\infty}^{\infty} \! \! dx \, \sech^2[s_-(x-p_i)] \tanh[s_-(x-p_i)] \left\{\tanh[s_+(x-p_{i+a_\alpha})]+a_\alpha\right\}^2.
\end{align}
The approximation in Eq.~(\ref{approx}) cannot be directly used as the integrand does not involve a $\sech[s_+(x-p_k)]$ factor as in Eq.~(\ref{form}). However, by integrating by parts a $\sech[s_+(x-p_k)]$ factor can be introduced to yield  
\begin{align}
    I_1(\text{even})=s_+ \int_{-\infty}^{\infty} \! \! dx \,
    \sech^2[s_+(x-p_{i+a_\alpha})] \left\{\tanh[s_+(x-p_{i +a_\alpha})]+a_\alpha\right\} \sech^2[s_-(x-p_i)].
\end{align}
We can now use the approximations given in Eqs.~(\ref{approx}) and (\ref{approx2}) as in the previous cases to obtain
\begin{equation}\label{eveni1}
    I_1(\text{even})\simeq \left(8-8 \frac{s_-}{s_+}\right) \times
    \begin{cases}
        e^{-2 s_- l_i}, &\alpha=+ \\
       - e^{-2 s_- l_{i-1}}, &\alpha=-. 
    \end{cases}
\end{equation}
The remaining integrals in Eq.~(\ref{rhsraw}) are evaluated for the case of even $i$ in the same way. The non-vanishing ones are listed below. 
\begin{align}\label{eveni2}
 \int_{-\infty}^{\infty} \! \! dx \, \Omegaid \omega_\alpha^3 \simeq
 -\left(16 - 24 \frac{s_-}{s_+}\right)\times
 \begin{cases}
    e^{-2 s_- l_i}, & \alpha=+ \\
    -e^{-2s_-l_{i-1}}, & \alpha=-,
 \end{cases}
\end{align}
\begin{align}\label{eveni3}
 \int_{-\infty}^{\infty} \! \! dx \, {\Omegaid}\omega_\alpha ({\omega_\alpha}_x+\Omegaid) \simeq
 -8 s_-\left(1-\frac{s_-}{s_+}\right)\times
 \begin{cases}
    e^{-2 s_- l_i}, & \alpha=+ \\
    -e^{-2s_-l_{i-1}}, & \alpha=-,
 \end{cases}
\end{align}
\begin{align}\label{eveni4}
\int_{-\infty}^{\infty} \! \! dx \, \Omegaid \Omegai {\omega_\alpha}_x \simeq
 8 s_-\times
 \begin{cases}
    e^{-2 s_- l_i}, & \alpha=+ \\
    -e^{-2s_-l_{i-1}}, & \alpha=-.
 \end{cases}
\end{align}
Using the results from Eqs.~(\ref{eveni1}), (\ref{eveni2}), (\ref{eveni3}), and (\ref{eveni4}) in Eq.~(\ref{rhsraw}) we obtain
\begin{equation}\label{rhseven}
    \text{RHS (even)} \simeq \left\{8-16 \e s_-\left(1+\frac{s_-}{s_+}\right) \right\}e^{-2s_-l_{i-1}}
    - \left\{8-16 \e s_-\left(1+\frac{s_-}{s_+}\right) \right\}e^{-2 s_-l_i}.
\end{equation}
Finally, Eqs.~(\ref{lhsodd}) and (\ref{rhsodd}) are put together to yield
\begin{align}
     -\frac{4}{3}s_+ \dot{p}_i
    + 8 s_- \dot{p}_{i+1} e^{-2 s_- l_i} &
     + 8 s_- \dot{p}_{i-1} e^{-2 s_- l_{i-1}} \nonumber \\
     &=16 \e \left( \frac{s_+}{3}+2 s_- \frac{s_-}{s_+} \right)e^{-2 s_- l_{i-1}}
    -16 \e \left( \frac{s_+}{3}+2 s_- \frac{s_-}{s_+} \right)e^{-2 s_- l_{i}},
\end{align}
where $i$ is odd. Similarly, Eqs.~(\ref{lhseven}) and (\ref{rhseven})  yields
\begin{align}
 -\frac{4}{3}s_- \dot{p}_i+8 s_- \dot{p}_{i+1}e^{-2 s_-l_i} &+8 s_- \dot{p}_{i-1}e^{-2 s_-l_{i-1}} \nonumber \\
 &=\left\{8-16 \e s_-\left(1+\frac{s_-}{s_+}\right) \right\}e^{-2s_-l_{i-1}}
    - \left\{8-16 \e s_-\left(1+\frac{s_-}{s_+}\right) \right\}e^{-2 s_-l_i},
\end{align}
where $i$ is even. 

\section{Weakly non-linear analysis}\label{appendix:landau}
This appendix is dedicated to the analysis of the primary bifurcation point of the stationary solutions of the dAC equation. In Sec.~\ref{sec:oneperiod}, we identified the point $L=L_c$ as the primary bifurcation point when the domain size $L$ was varied as the continuation parameter keeping the driving strength $\e$ fixed. Here we perform a Landau-Stuart type analysis to derive a dynamical equation for the amplitude of the first unstable mode near $L_c$, and then by analysing the fixed points of this equation we establish that the primary bifurcation point is super critical for any value of $\e$. Note that a similar analysis was performed for the case of the cCH equation in Ref.~\citep{Tseluiko_2020}.  

Proceeding along similar lines we set $k=k_c-\epsilon^2$, where $\epsilon$ is a small parameter and $k=2 \pi/L$, and introduce the scaled coordinate $\xi=k x$ and the slow time scale $\tau=\epsilon^2 k t$. Note that $k_c=2 \pi/L_c=1$. The dAC equation is written in terms of the scaled variables as
\begin{equation}\label{rescaleddac}
    \epsilon^2 k \psi_\tau = k^2  \psi_{\xi \xi} + \psi-\psi^3 + 2 k \e \psi \psi_{\xi}.
\end{equation}
We now expand $\psi$ as a series in $\epsilon$: 
\begin{equation}\label{seriesineps}
    \psi=\psi_0+\epsilon \psi^{(1)}(\xi,\tau)+\epsilon^2 \psi^{(2)}(\xi,\tau)+...,
\end{equation}
where $\psi_0=0$, as we are perturbing about this solution near the primary bifurcation point. Substituting Eq.~(\ref{seriesineps}) in Eq.~(\ref{rescaleddac}) and collecting terms of order $O(\epsilon)$ we get

\begin{equation}
     \psi^{(1)}_{\xi \xi}+\psi^{(1)}=0.
\end{equation}
 The above equation is readily solved with periodic boundary conditions to yield
\begin{equation}
    \psi^{(1)}=A_1 (\tau) e^{i \xi}+ A_1^* (\tau) e^{-i \xi}.
\end{equation}
 We proceed to higher order terms in $\epsilon$ to obtain the time dependent coefficient $A_1(\tau)$. At order $O(\epsilon^2)$ we get
\begin{align}
    \psi^{(2)}_{\xi \xi}+\psi^{(2)}&=-2 i \e A_1^2 e^{2 i \xi}+c.c. \nonumber \\  
    \implies \psi^{(2)}&=A_2 (\tau) e^{i\xi}+\left(i \frac{2}{3}\e A_1^2 \right)  e^{2 i \xi}+c.c.
\end{align}
Similarly, at order $O(\epsilon^3)$ we get
\begin{equation}
    \psi^{(3)}_{\xi \xi}+\psi^{(3)}=C_1 e^{i \xi}+C_2 e^{2i\xi}+C_3 e^{3 i \xi}+c.c.,
\end{equation}
where the coefficient
\begin{equation}
    C_1={A_1}_\tau-2A_1+\left(3+\frac{4}{3} \e^2 \right) A_1^2 A_1^*,
\end{equation}
and the coefficients $C_2$ and $C_3$ are functions of $A_1$ and $A_2$. Their explicit forms are not relevant to this analysis and are therefore not shown. Now we set $C_1=0$ to avoid secular terms yielding 
\begin{equation}
   {A_1}_\tau=2A_1-\left(3+\frac{4}{3} \e^2 \right) A_1^2 A_1^*. 
\end{equation}
It follows that
\begin{equation}
 \frac{d |A_1|}{d \tau}= \left[ 2-\left( \frac{4 \e^2}{3}+3\right) |A_1|^2\right]|A_1|.   
\end{equation}
The above equation has an unstable fixed point at $|A_1|=0$ and a stable fixed point at $|A_1|=\sqrt{6/(4\e^2+9)}$. This implies that there exists stable small amplitude harmonic solution near the primary bifurcation point. Also, the stability of the fixed points do not depend on the value of the driving strength $\e$. Hence, we conclude that the primary bifurcation point is super critical for all values of $\e.$  
\end{widetext}

\bibliography{references}
\end{document}